\documentclass[twocolumn,showpacs,aps,prd,floatfix]{revtex4-1}
\usepackage{graphicx}
\newcommand{\beq}{\begin{equation}}
\newcommand{\eeq}{\end{equation}}
\newcommand{\beqn}{\begin{eqnarray}}
\newcommand{\eeqn}{\end{eqnarray}}

\newcommand{\mdisk}{M_{disk}}
\newcommand{\rdisk}{R_{disk}}
\newcommand{\rin}{R_{in}}
\newcommand{\tdisk}{t_{disk}}
\newcommand{\torb}{t_{orb}}
\newcommand{\tvis}{t_{vis}}
\newcommand{\tgw}{t_{merge}}

\newcommand{\Ls}{L_{syn}}
\newcommand{\Lb}{L_{brem}}

\newcommand{\aj}{Astron.\ J.\ Supp.\ }

\newcommand{\apjl}{Astrophys.\ J.\ Lett.\ }
\newcommand{\mnras}{Mon.\ Not.\ R.\ Astron.\ Soc.\ }

\newcommand{\aap}{Astron.\ and\ Astrophys.\ }

\newcommand{\pasj}{PASJ}

\usepackage{dcolumn}
\usepackage{bm}
\usepackage{epsf}

\begin{document}
\bibliographystyle{apsrev4-1}
\title{Binary black hole mergers in gaseous
  disks: Simulations in general relativity}

\author{Brian D. Farris}

\author{Yuk Tung Liu}

\author{Stuart L.\ Shapiro}

\altaffiliation{Also at Department of Astronomy \& NCSA, University of Illinois
at Urbana-Champaign, Urbana, Illinois 61801}

\affiliation{Department of Physics, University of Illinois at
Urbana-Champaign, Urbana, Illinois~61801}

\begin{abstract}
Simultaneous gravitational and electromagnetic 
wave observations of merging black hole binaries (BHBHs)
can provide unique opportunities to study gravitation physics, 
accretion and cosmology.  Here we perform fully general-relativistic,
hydrodynamic simulations of equal-mass, nonspinning BHBHs
coalescing in a circumbinary disk. We evolve the metric 
using the Baumgarte-Shapiro-Shibata-Nakamura (BSSN) formulation of 
Einstein's field equations with
standard moving puncture gauge conditions. We handle the hydrodynamics
via a high-resolution shock-capturing scheme. These initial simulations
are exploratory in nature and simplified accordingly. We track the inspiral starting from a binary separation of $10 M$, where $M$ is the total binary mass. We take the disks
to have an inner radius at $\rin \approx 15M$ to account for the
hollow created by the binary torques.  Our disks extend to $R\approx 65M$
and have an initial scale height of $H/R \approx 0.03-0.11$.  The gas
is governed by a $\Gamma$-law equation of state, with $\Gamma$ equal to $5/3$, $4/3$, and $1.1$. Disks are allowed to relax
in the ``early inspiral'' epoch to provide quasistationary
realistic initial data. We then evolve the spacetime metric and matter during the 
``late inspiral and merger'' epochs.  The later simulations are designed to 
track BHBH inspiral following disk-binary decoupling, through merger and 
ringdown, terminating before viscosity has time to fill the hollow about the black hole remnant. We compute the gas flow and accretion rate 
and estimate the electromagnetic
luminosity due to bremsstrahlung and synchrotron emission as a perturbation for optically
thin disks. The  synchrotron component of the luminosity peaks in the infrared band and should be 
detectable by WFIRST and possibly the LSST for a $10^8 M_{\odot}$ binary embedded in a disk with a 
density $n \sim 10^{12} \mbox{ cm}^{-3}$ at $z=1$, beginning with a maximum value
of $L\sim 10^{46} n_{12}^2 M_8^3 \mbox{erg } \mbox{s}^{-1}$ at decoupling, and decreasing steadily over a timescale of $\sim 100 \ M_8$ hours to a value of $L\sim 10^{45} n_{12}^2 M_8^3 \mbox{erg } \mbox{s}^{-1}$ at merger. 
\end{abstract}

\maketitle

\section{Introduction}
All bulge galaxies (including the Milky Way) are believed to contain
a central supermassive black hole (SMBH) with a mass $M$ between $10^4 M_{\odot}$ and  $10^9
M_{\odot}$ \cite{richstone98,peterson00,ferrarese05}.  It is also believed that galaxy mergers commonly lead to the formation of a
massive black hole binary (BHBH) system in the merged remnants \cite{begelman80,roos81}.  In
the standard picture, the BHBH binary separation decreases, first through dynamical
friction due to distant stellar encounters, then through gravitational slingshot interactions in which
nearby stars are ejected at velocities comparable to the binary's orbital velocity, and
finally through the emission of gravitational radiation, leading to
coalescence of the binary \cite{merritt05}.
These low-frequency gravitational waves will be detectable by LISA
and will contain a wealth of information about the inspiral.  The gaseous accretion flow that forms around
the binary can be a source of electromagnetic radiation as well.
Following the detection of gravitational waves from a BHBH merger,
electromagnetic ``afterglow'' radiation can provide confirmation of
the coalescence
\cite{milosavljevic05,rossi09,schnittman08,corrales10,oneill09,shapiro10,tanaka10}.
The timescale during which detectable ``afterglow'' radiation rises to its maximum is governed by viscous diffusion of gas close to the remnant and ranges from several years to hundreds of decades in the case of
supermassive BHBH systems.

There also exists the possibility of detecting electromagnetic
``precursor'' radiation prior to the merger and before the maximum
gravitational wave emission \cite{armitage02,chang10}.  If the merger takes place in a hot gas cloud in which the distant
gas is nearly homogeneous and either at rest with respect to the
binary (``binary Bondi'' accretion) or moving (``binary
Bondi-Hoyle-Lyttleton'' accretion), then Farris {\it et al}. \cite{farris10} (hereafter Paper I) have shown that the luminosity
will peak at the end of the binary inspiral phase immediately prior to
the final plunge.  At this stage shock heating of the
gas and turbulent magnetic field amplification are strongest.  The
peak luminosity lasts for $\delta t \sim M_6$ hours prior to merger
and then plummets sharply following the coalescence.  Here $M_6$ is
the binary mass in units of $10^6 M_{\odot}$.  If, instead, the
accretion takes place via a geometrically-thin, optically-thick
Keplerian disk around the binary (``binary Shakura-Sunyaev''
accretion), there may be a late-time precursor brightening from tidal
and viscous (or turbulent magnetic) dissipation in the inner disk.
This radiation peaks on a timescale of $\delta t \sim 0.1 M_6$ days
prior to merger and remains high afterwards \cite{chang10}.  Each
of these scenarios raises the exciting possibility of a
simultaneous detection of electromagnetic and gravitational waves from a BHBH merger.  

This picture is loosely supported by a number of observed AGNs that may be harboring BHBH binaries.  Very-long baseline interferometry (VLBI) observations of the elliptical galaxy 0402+379 have discovered two radio sources at a projected separation of only 7 pc.  The existence of jets, as well as variability associated with BH activity, indicate that the system may be a BHBH binary \cite{maness04,rodriguez06,rodriguez09}.  Another candidate is OJ 287, a BL Lac object whose light curve shows variability with a period of $\sim 12$~yr.  It is believed that this may be a massive BHBH binary around which the smaller BH orbits with a period of 12~yr, penetrating the disk of the primary and giving rise to the observed variability \cite{lehto96,valtonen06,valtonen08}. It has been proposed that the quasar SDSS 092712.65+294344 may be either a binary system \cite{bogdanovic09,dotti09}, or a recoiling BH that is the product of a binary merger \cite{komossa08}. Such suggestions are supported by a systematic shift of $2650~\mbox{km s}^{-1}$ in the emission lines.  Another candidate is quasar SDSS J153636.22+044127.0, in which two broad-line emission systems are observed, separated in velocity by $3500~\mbox{km s}^{-1}$.  This observation has been interpreted as a BHBH binary system in which each object has its own emission system \cite{boroson09}. Recently, the first triple AGN system, SDSS J1027+1749, has been discovered \cite{liushenstrauss11}. This galaxy contains three emission-line nuclei corresponding to a supermassive black hole triple with kpc-scale separations.

Information from a simultaneous detection of electromagnetic and gravitational waves may be useful for studying fundamental aspects of gravitational physics.  For example, in some modified gravity scenarios, the propagation velocity for gravitons may differ from that of photons \cite{kocsis08,deffayet07}.  Additionally,
the measurement of the luminosity distance from the gravitational wave
signal at an accuracy of $1\%-10\%$, coupled with the redshift
information from the electromagnetic detection, could serve as a
cosmological ``standard siren'' of unprecedented accuracy (better than
$\sim 1 \%$) \cite{holz05}.  Such detections may also combine accurate
measurements of BH spins and masses obtained from gravitational wave
signals with electromagnetic observations to probe BH accretion
physics in great detail \cite{kocsis06}.  It has even been proposed
that simultaneous detections of electromagnetic and gravitational
waves may provide a means of witnessing the birth of a quasar \cite{tanaka10b}.

Several mechanisms for electromagnetic emission from accretion disks
around merging BHBH binaries have been proposed.  In one
scenario, the inner edge of the accretion disk is identified as the radius
at which the viscous torque on the gas balances the gravitational torque from
binary.  This radius is between 1.5 and 2 times the orbital separation \cite{artymowicz94,gunther02,escala05,macfadyen08} and encompasses a hollow region in the disk.  Late in the inspiral the BHBH binary decouples from the disk and coalesces.  For a binary of mass $M \approx 10^6 M_{\odot}$ accreting at $10\%$ of the Eddington rate, the subsequent evolution of this disk, which is optically thick, gives rise to a source that initially peaks in the {\it UV} band and then hardens to extreme ultraviolet and soft x-ray emission at late times \cite{milosavljevic05,shapiro10,tanaka10}.  
Additionally, there is a sudden
change in the mass of the binary during the final stage of the merger,
as gravitational waves carry away a few percent of the mass.  The
abrupt change in the potential creates waves in the disk which may
grow into shocks and increase the luminosity of the disk in a
unique way \cite{corrales10,oneill09,megevand09}, giving rise to a detectable prompt x-ray signal.  Another possibility is that the
merged BH remnant may experience a recoil velocity which can, in principle, be as high as several thousand km
$\mathrm{s}^{-1}$ \cite{campanelli07}, although it is likely to be much lower ($< 200 \ \mbox{km}/\mbox{s}$) in most galaxy mergers \cite{bogdanovic07}.  This recoiling BH may ``shake'' or penetrate
the disk, creating shocks which could give rise to a transient
signal. 

Various methods have been used to model plausible sources of electromagnetic emission from BH mergers.  In one approach, the dynamics of the inspiral is ignored, focusing on the effect of BH kicks and/or BH mass loss on the hydrodynamical flow \cite{anderson10,megevand09,schnittman08,shields08,lippai08,rossi09,oneill09,corrales10,zanotti10}.  In another approach, the behavior of the gas is modeled by following the motion of collisionless ``particle tracers'' on geodesics \cite{vanmeter09}.  Other approaches involve vacuum and/or force-free calculations to investigate the role that magnetic fields may play in producing detectable electromagnetic emission when the density near the binary at merger is very low \cite{mosta10,palenzuela10}. Only recently have fully relativistic, hydrodynamical simulations of BHBH binary inspiral and merger in a gaseous environment been performed \cite{farris10,bode10,bogdanovic10,bode11}.

In this paper we study BHBH binary mergers in the presence of a circumbinary gaseous disk.  Modeling such systems requires fully general-relativistic dynamical simulations.  The development of stable algorithms to integrate Einstein's field equations of general relativity numerically in $3+1$ dimensions, such as the Baumgarte-Shapiro-Shibata-Nakamura (BSSN) formalism~\cite{bssn_shibata,bssn_stu} and the generalized harmonic approach~\cite{friedrich85,garfinkle02,pretorius05a}, together with the identification of suitable gauge conditions, has enabled several pioneering simulations that demonstrate how to track the late inspiral, plunge and merger of a BHBH binary in vacuum~\cite{pretorius05b,campanelli06,baker06}.  More refined follow-up simulations of these strong-field, late phases, joined onto analytic, post-Newtonian calculations of the early inspiral epoch~\cite{blanchet08}, are now capable of producing accurate gravitational waveforms for merging BHBH binaries with companions spanning a range of mass ratios and spins (see e.g. \cite{ninja} and references therein).

With the problem of gravitational wave emission from a vacuum BHBH binary inspiral well in hand, it is now important to turn to the problem of electromagnetic emission from BHBH binary coalescence in an astrophysically realistic gaseous environment.  
 
In Paper I, we considered {\it hot} accretion flows in which the gas
is near the galaxy virial temperature and the specific angular
momentum of the gas $\tilde{L}$ is less than that of a circular orbit
near the horizon, $\tilde{L} \lesssim M c$.  Such flows can be well described by the
spherical Bondi, or Bondi-Hoyle-Lyttleton accretion model.  In this paper, we consider flows in which
the angular momentum of the gas cannot be neglected, and the accretion
flow is disklike.  

Disk accretion onto a BHBH binary has been studied previously 
in the Newtonian, geometrically thin-disk limit, both analytically \cite{larwood97,ivanov99,armitage02,shapiro10,chang10,liu10} and numerically \cite{rossi09,corrales10,oneill09,macfadyen08}. 
We extend this work by performing fully relativistic  
hydrodynamcal simulations in 3+1 dimensions. Our treatment
here is quite preliminary and meant to introduce the computational 
framework for  more detailed and 
realistic simulations that we are preparing. In this paper we restrict our
attention to a circumbinary disk residing in the orbital plane of 
two nonspinning, equal-mass, binary black holes. The black holes are 
initially in a quasistationary, 
nearly circular orbit and represent a solution to the conformal thin-sandwich
(CTS) initial-value equations (see, e.g. \cite{pfeiffer03,cook04,baumgartebook10} and references therein). The mass of the disk is assumed to be
small in comparison to the total black hole mass. We explore the response of
the disk to the binary on a {\it dynamical} timescale and thus ignore 
the secular motion of gas in the disk due to viscosity or turbulent 
magnetic fields.  We treat the gas as a perfect fluid described by a 
$\Gamma$-law equation of state (EOS) and handle 
shocks by high-resolution, shock-capturing (HRSC) techniques.
We study the response of the disk to tidal torques during 
the early and late inspiral phases, as well as during the merger and post-merger
epochs. The inspiral and merger are followed by solving the BSSN field equations
\cite{bssn_shibata,bssn_stu} with moving puncture gauge conditions \cite{baker06,campanelli06}. We are particularly
interested in the evolution of the 
hollow region in the disk about the binary \cite{artymowicz94,gunther02,escala05,macfadyen08}
and the extent to which gas penetrates the hollow
and accretes onto the black holes. We also estimate, as a perturbation, 
the electromagnetic emission from the disk that characterizes 
the inspiral and merger epochs.
Our treatment is appropriate for describing the  
epoch following disk-binary ``decoupling,'' when the BHBH inspiral timescale is 
much shorter than the viscous timescale in the disk, whereby viscosity-induced
inflow can be neglected. Our analysis remains 
valid throughout the binary merger and ringdown phases, but is no longer 
adequate to describe the late-time evolution when viscosity serves to 
drive gas into the hollow and accrete onto the merged 
remnant \cite{milosavljevic05,shapiro10}.  
We  briefly compare our results 
with another, recently published, general relativistic study \cite{bode11} that treats a similar 
scenario, but employs different methods (e.g. different initial data and luminosity estimates) and addresses 
different issues.

The structure of the paper is as follows.  In Sec.~\ref{sec:computational_challenge} we summarize the computational challenge posed by the large dynamic range associated with this problem, and we describe our approach for overcoming this challenge.  In Secs.~\ref{sec:basic_eqns} and \ref{sec:numerical} we briefly outline the basic gravitational field and matter evolution equations and their specific implementations in our numerical scheme.  Here we also provide an overview of our initial data, gauge conditions, and diagnostics.  In Sec.~\ref{sec:code_tests}, we review code tests that were performed to validate our numerical scheme.  In Sec.~\ref{sec:results} we describe the results of our binary BHBH merger simulations.  In Sec.~\ref{sec:discussion} we summarize our findings, and briefly compare with previous simulations. Henceforth we adopt geometrized units and set $G=c=1$.
\section{Computational Challenge}
\label{sec:computational_challenge}
Simulating realistic accretion flows is extremely challenging due to
the enormous dynamic range characterizing the time and length scales in the
problem.  One length scale is set by the initial, total Arnowitt-Deser-Misner (ADM) mass of the
binary system, $M$.  We neglect the mass of the
disk and assume that $\mdisk \ll M$.  The ADM mass sets the length scale at
which relativistic effects become significant.  Another important
length scale is the binary orbital separation $a$.  Associated with
the orbital separation is the orbital period, $\torb \equiv 2 \pi/\Omega_{bin}\approx 2 \pi
(a/M)^{3/2}$, where $\Omega_{bin}$ is the binary orbital angular velocity.

Torques from the
binary have the effect of driving matter in the vicinity of the BHBH orbit
outward, creating a hollow cavity inside the inner edge of the
accretion disk at $\rin \sim 1.5-2 a$ \cite{artymowicz94,gunther02,escala05,macfadyen08}.  This radius is determined by the balance between
viscous stresses in the disk and tidal torques from the binary
\cite{macfadyen08,artymowicz94}, provided that the viscous
timescale $\tvis \approx (2/3) \rin^2/\nu$ is shorter than the gravitational
inspiral timescale $\tgw \approx (5/16) a^4 / M^3$.  Here $\nu$ is the
shear viscosity, and we assume an equal-mass binary.  As the orbital
separation decreases, the binary eventually enters an epoch at which
$\tgw < \tvis$.  At this point, the binary decouples from the disk
\cite{milosavljevic05,armitage02,liu03,liu10}.  If one assumes an
$\alpha$-disk with a viscosity law $\nu(R) = (2/3)\alpha P_{gas} /
(\rho \Omega_K)$, where $\rho$ is the gas density and $P_{gas}$ is the
gas pressure, then the decoupling radius $a_d$ is given by
\cite{milosavljevic05,tanaka10,liu10}
\begin{equation}
\label{eq:ad}
\frac{a_{\rm d}}{M}  \approx 126  \alpha_{-1}^{-17/50} S^{-49/200} 
\lambda ^{7/10} M_6^{2/25}(\delta_{-1} \zeta)^{17/40} \theta_{0.2}^{-17/200}\ ,
\end{equation}
where $\alpha = 0.1 \alpha_{-1}$, $\delta = 0.1 \delta_{-1}$,
$S \equiv 3 \pi \Sigma(a_{\rm d}) \nu(a_{\rm d})/{\dot M}_{\rm Edd}$ and 
$\theta = 0.2 \theta_{0.2}$. Here ${\dot M}_{\rm Edd} = 4 \pi M m_p/(\eta 
\sigma_T)$ is the Eddington
accretion rate, $\sigma_T$ is the Thomson cross section for electron 
scattering, $\eta$ is the radiative efficiency, $\theta < 1$ is a porosity 
correction factor applied to the 
scattering-dominated optical depth ~\cite{turner04}, and $\delta$
roughly accounts for the shortening of the viscous timescale at the
disk edge where the surface density $\Sigma$ is very steep
\cite{lyndenbell74}.

Another important
length scale is the characteristic size of the disk, $\rdisk$, which
we define here as the radius at which the gas pressure is maximum,
$\rdisk \equiv R(P_{max})$.  In general, $\rdisk$ depends on the
details of the temperature and angular momentum profile in the disk, and is
highly dependent on the particular choice of disk model.  Associated
with $\rdisk$ is the orbital time scale $\tdisk$ which we define as the
Keplerian orbital period $\tdisk = 2 \pi (\rdisk / M)^{3/2}$.

If we assume the size of the entire disk is $\mbox{several }
\rdisk \gg \rin$ and use the estimate of $a_{\rm d}$ given in
Eq.~(\ref{eq:ad}), we find that a simulation of the full
inspiral from decoupling to
merger would require us to resolve length scales from $\sim M$ to
$\gtrsim 10^3 M$. More challenging, we must resolve timescales from $M$ to
$\mbox{several } \tgw \sim 10^8 M$.
Unfortunately, the latter is beyond the capability of current numerical
codes.  In order to circumvent this issue, we consider a disk with
relatively small values of $a_d = 10M$, $\rin \sim 15 M$, and $\rdisk
\sim 35 M$. With these choices, the important time scales become
$\torb=225 M$, $\tdisk =1300 M$,  and $\tgw=1250M$.
Given the wide range of gaseous
environments in galactic cores, such parameters are not implausible, and
we expect that qualitative features of our results can be extended to
disks with larger values of $a_d$, $\rin$, and
$\rdisk$.  Our choice allows us to study the full evolution of the system
from decoupling to binary merger.  We focus on the post-decoupling phase through merger, ringdown, and disk equilibration, but prior to disk inflow on viscous
time scales. Accordingly, our
perfect-fluid approximation will closely mimic a realistic
flow during these epochs, as the viscous time scale (which may originate from MHD
turbulence) is long compared to the
length of our simulations. 
\section{Basic Equations}
\label{sec:basic_eqns}
Throughout this paper, we use Latin indices to denote spatial components (1-3) and Greek indices to denote spacetime components (0-3). 

\subsection{Early inspiral epoch}
\label{sec:early_epoch}
We define the ``early inspiral epoch'' as the phase of the binary inspiral prior to decoupling. Throughout this phase, the inspiral time scale is much longer than the orbital time scale. This fact can be exploited by neglecting the change in binary separation and employing a metric that is quasistationary in the rotating frame of the binary. This simplification provides an accurate solution for the spacetime without the computational expense of a full evolution of Einstein's field equations. We evolve the full relativistic hydrodynamics equations in this background metric over $\sim 5 \tdisk$ to enable the disk to relax to a quasistationary state. This technique thus provides astrophysically realistic initial data with which to begin evolution of the late inspiral and merger epochs (Sec.~\ref{sec:late_epoch}).

In order to use this method, we must choose a coordinate system in which the metric explicitly reflects the symmetry of the spacetime. This symmetry, describing a
spacetime that is quasistationary in a frame that rotates with the
orbital frequency of the binary $\Omega$, can be constructed by employing a
helical Killing vector,
\begin{equation}
  \label{killing_vec}
  \xi \equiv \partial_t + \Omega \partial_{\phi}. 
\end{equation}
For a spacetime admitting such a Killing
vector, we have 
\begin{equation}
  \label{killing_eq}
  {\cal L}_{\xi} g_{\mu \nu}=0 \ ,
\end{equation}
where ${\cal L}$ is the Lie derviative, and
$g_{\mu \nu}$ is the spacetime metric.

Provided we are working in an appropriate coordinate system (i.e. one
employing Killing coordinates $t$ and $\phi$), we may express the
metric at any point in spacetime in terms of the
metric on an initial $t=0$ slice according to
\begin{equation}
\label{gab_rotate}
g_{\mu \nu}(t,r,\theta,\phi)\doteq g_{\mu \nu}(0,r,\theta,\phi-\Omega t)
\end{equation}
where the symbol $\doteq$ denotes that the equality holds only in a particular 
coordinate system. One can easily verify that the above equation satisfies Killing's equation~(\ref{killing_eq}). 

We note that Eq.~(\ref{gab_rotate}) is written in spherical polar
coordinates, i.e. $\{x^{\alpha}\}= \{t,r,\theta,\phi\}$. However, Cartesian coordinates are more suitable for work in 3D, as coordinate singularities at $r=0$ and on the polar axis are avoided. We therefore transform the
spherical components of $g_{\mu \nu}$ back to the Cartesian components using the usual tensor transformation formula.

BHBH evolution employing standard puncture initial data and 
moving puncture gauge conditions does not 
result in a metric that satisfies Eq~(\ref{killing_eq}) (Puncture initial data does not implement a helical Killing vector). By contrast,
BHBH CTS initial data (see, e.g. \cite{cook04}) are
specifically constructed to satisfy this equation: CTS initial data impose the condition that the spacetime in the rotating frame is
stationary (see \cite{baumgartebook10} for discussion and references).
This condition is valid, approximately, whenever the binary companions are sufficiently
well separated that the inspiral time scale is much longer 
than the orbital time scale.  In this quasistationary 
early inspiral regime we can employ CTS 
initial data and CTS lapse and shift functions to evolve the metric via a simple coordinate rotation in lieu of integrating the Einstein field equations. 
We can then evolve the disk by integrating the hydrodynamic equations for the
fluid in this background spacetime.

CTS initial data contains excised interiors. We follow the
technique outlined in \cite{etienne07} and fill the excised region inside the BH interiors with
smoothly extrapolated ``junk'' (i.e., constraint-violating) data. This
treatment is valid because the interior regions are causally
disconnected from the exterior spacetime.

\subsection{Late inspiral and merger epochs}
\label{sec:late_epoch}
We evolve both the metric and hydrodynamic equations during the late inspiral and merger epochs. Our basic gravitational field and relativistic hydrodynamics equations
are discussed in \cite{bssn_stu,mhd_code_paper}, where their numerical
implementation is described and detailed code tests are summarized.  Here, we
briefly review these equations and their implementation.

We write the spacetime metric in the standard $3+1$ form,
\beq
ds^2 = - \alpha^2 dt^2 + \gamma_{ij}(dx^i + \beta^i dt)(dx^j + \beta^j dt) .
\eeq
where $\alpha$, $\beta^i$, and $\gamma_{ij}$ are the lapse, shift, and
spatial metric, respectively.  The extrinsic curvature $K_{ij}$ is
given by
\begin{equation}
\label{Kij}
(\partial_t - {\mathcal{L}}_{\beta})\gamma_{ij} = -2\alpha K_{ij},
\end{equation}
where ${\mathcal{L}}_{\beta}$ is the Lie derivative with respect to
$\beta^i$.  We evolve $\gamma_{ij}$ and $K_{ij}$
using the BSSN formulation~\cite{bssn_shibata,bssn_stu}.  The fundamental variables for
BSSN evolution are
\begin{eqnarray}
  \label{phidef}
  \phi &\equiv& \frac{1}{12}\ln[\det(\gamma_{ij})]\ , \\
  \tilde\gamma_{ij} &\equiv& e^{-4\phi}\gamma_{ij}\ , \\
  K &\equiv& \gamma^{ij}K_{ij}\ , \\
  \tilde A_{ij} &\equiv& e^{-4\phi}(K_{ij} - \frac{1}{3}\gamma_{ij}K)\ , \\
  \tilde\Gamma^i &\equiv& -\tilde\gamma^{ij}{}_{,j}\ .
\end{eqnarray}
The evolution and constraint equations for
these fields are summarized in~\cite{bssn_shibata,bssn_stu}.  We assume in this paper that the mass of the gas is negligible compared to the mass of the BHs, thus we do not include matter source terms in our metric evolution equations.

Adding Kreiss-Oliger dissipation to the BSSN evolution equations outside the BH can reduce high-frequency numerical noise associated with adaptive mesh refinement (AMR) refinement interfaces \cite{baker06}.  We use this technique here and have found it useful in reducing Hamiltonian and momentum constraint violations.

We adopt the standard moving puncture gauge conditions: an advective ``1+log'' slicing condition for the lapse and a ``Gamma-freezing'' condition for the shift \cite{van_meter_06}.  Thus we have
\beqn
\partial_0 \alpha &=& - 2 \alpha K ,\\
\partial_0 \beta^i &=& (3/4) B^i , \\
\partial_0 B^i &=& \partial_0 \tilde{\Gamma}^i - \eta B^i \ , 
\eeqn
where $\partial_0 = \partial_t - \beta^j \partial_j$.  The $\eta$ parameter is set to $0.5/M$ in all simulations.

\subsection{Evolution of hydrodynamic fields}

The fundamental matter variables are the rest-mass density $\rho_0$, specific internal energy $\epsilon$, pressure $P$, and four-velocity $u^{\mu}$.  The stress-energy tensor for an ideal gas is given by
\[
T_{\mu \nu} = \rho_0 h u_{\mu} u_{\nu} + P g_{\mu \nu} \ ,
\]
where $h=1+\epsilon + P/\rho_0$ is the specific enthalpy.  We evolve the ``conservative'' variables $\rho_*$, $\tilde{S}_i$, and $\tilde{\tau}$.  They are defined as
\beqn
\rho_* &=& - \sqrt{\gamma} \rho_0 n_{\mu} u^{\mu}\\
\tilde{S}_i &=& \sqrt{\gamma} T_{\mu \nu} n^{\mu} \gamma^{\nu}{}_i\\
\tilde{\tau} &=& \sqrt{\gamma} T_{\mu \nu} n^{\mu} n^{\nu} - \rho_* \ .
\eeqn
Here $n^{\alpha} = (\alpha^{-1},-\alpha^{-1}\beta^i)$ is the timelike unit vector normal to the $t=\mbox{constant}$ time slices.
Evolution equations are given by Eqs. (34), (36), and (38) of \cite{mhd_code_paper}:
\beqn
\label{hydro_evolution_rhostar}
\partial_t \rho_* + \partial_j ( \rho_* v^j) &=& 0, \ \ \ \ \\\
\label{hydro_evolution_Stilde}
\partial_t \tilde{S}_i + \partial_j (\alpha \sqrt{\gamma} T^j{}_i) &=& \frac{1}{2} \alpha \sqrt{\gamma} T^{\alpha \beta} \partial_i g_{\alpha \beta},\\
\label{hydro_evolution_tau}
\partial_t \tilde{\tau} + \partial_i ( \alpha^2 \sqrt{\gamma} T^{0i} - \rho_* v^i) &=& s \ ,
\eeqn
where $\gamma \equiv \mbox{det}(\gamma_{ij}) = e^{12 \phi}$ and $v^i \equiv u^i/u^0$ is the fluid's 3-velocity.  The energy source term $s$ is given by 
\beqn
s &=& -\alpha \sqrt{\gamma} T^{\mu \nu} \nabla_{\nu} n_{\mu} \nonumber\\
&=& \alpha \sqrt{\gamma} [ ( T^{00} \beta^i \beta^j + 2 T^{0i} \beta^j + T^{ij}) K_{ij} \nonumber \\
&& -(T^{00} \beta^i + T^{0i}) \partial_i \alpha] \ .
\eeqn

\subsection{Equation of state}
To complete the system of equations, we must specify an EOS.  While our code can handle any EOS of the form $P=P(\rho_0,\epsilon)$, we adopt a $\Gamma$-law EOS,
\begin{equation}
  \label{EOS}
  P=(\Gamma-1)\rho_0 \epsilon .
\end{equation}
We perform simulations with $\Gamma=4/3$, $5/3$, and $1.1$. By varying $\Gamma$ we effectively examine gas flow under a full range of conditions. We choose $\Gamma=5/3$ as our canonical case. The choice of $\Gamma=1.1$ approximates an isothermal gas (we have chosen $\Gamma=1.1$ rather than $\Gamma=1$ in order to retain the  $\Gamma$-law form
of the EOS while still approximating isothermality). At $t=0$, we take
the EOS to be isentropic, with $P=K\rho_0^{\Gamma}$, where
$K=\mbox{constant}$. Throughout this paper, we define temperature by 
\begin{equation}
  P= 2 n kT \ ,
\end{equation}
appropriate for pure ionized hydrogen.
\section{Numerical Methods}
\label{sec:numerical}
\subsection{Disk initial data}
\label{sec:initial_data}
 For our disk initial data, we use the equilibrium solution for a
 stationary disk around a {\it single} Kerr
BH derived by Chakrabarti {\it et al.} \cite{chakrabarti85} and
summarized in \cite{devilliers03}.  We take this disk as inital data
for a binary BHBH, placing the inner
boundary of the disk well outside the BHBH orbital radius.  Though no
longer stationary, the initial disk settles down to quasistationary
equilibrium as the binary rotates, apart from low amplitude spiral density
waves induced by the time-varying tidal torque.  For completeness, we provide a brief summary of the construction of disk initial data in Appendix~\ref{disk_id_appendix}. 

For our fiducial equation of state, $\Gamma=5/3$, the resulting outer disk
radius is $R_{out}\approx 65 M$ and the disk scale height at $\rdisk$
is $H/R = 0.11$ (see Table~\ref{table:unmag} for more details).  Here
$H$ is defined as the height above the equatorial plane where the
pressure falls to $1/e$ its value on the equatorial plane at the
radius of maximum pressure.  For binary BHs, the disk is approximately
stationary if $\rin \gg a$. Initially, we take $R_{in}/a = 1.5$. We find that the disk relaxes to a near
quasistationary state after a time $\sim 4 t_{disk}$.

\subsection{Evolution of metric and matter}
We evolve the BSSN field equations
with fourth-order accurate, centered, finite-difference stencils,
except on shift advection terms, where we use fourth-order accurate
upwind stencils.  We apply Sommerfeld outgoing wave boundary
conditions to all BSSN fields.  Our code is embedded in
the Cactus parallelization framework~\cite{Cactus}, and our
fourth-order Runge-Kutta time-stepping is managed by the {\tt MoL}
(Method of Lines) thorn, with a Courant-Friedrichs-Lewy factor
set to 0.5 in all BHBH simulations.  We use the
Carpet~\cite{schnetter04} infrastructure to implement the moving-box
AMR. In all AMR simulations presented here, we
use second-order temporal prolongation, coupled with fifth-order
spatial prolongation. The apparent horizon of the
BH is computed with the {\tt AHFinderDirect} Cactus
thorn~\cite{thornburg04}.

We write the general-relativistic hydrodynamics equations in
conservative form. They are evolved by an HRSC technique~\cite{mhd_code_paper} that employs the
piecewise parabolic (PPM) reconstruction scheme~\cite{colella84} coupled to
the Harten, Lax, and van Leer (HLL) approximate Riemann solver~\cite{HLL}. 
The adopted hydrodynamic scheme is second-order accurate for smooth 
flows, and first-order accurate when discontinuities (e.g.\ shocks) 
arise.  Throughout the
evolution, we impose limits on the pressure to prevent
spurious heating and negative values of the internal energy
$\epsilon$. Specifically, we require $P_{\rm min}\leq P \leq P_{\rm
  max}$ inside the horizon, 
where $P_{\rm max}=10 K \rho_0^\Gamma$ and $P_{\rm min}=K 
\rho_0^\Gamma/2$. 
Whenever $P$ exceeds $P_{\rm max}$ or drops below $P_{\rm min}$, we 
reset $P$ to $P_{\rm max}$ or $P_{\rm min}$, respectively.  We check that this fix is applied only inside the apparent horizon, which is causally disconnected from the rest of the grid.

At each timestep, 
we need to recover the ``primitive variables'' 
$\rho_0$, $P$, and $v^i$ from the ``conservative'' variables 
$\rho_*$, $\tilde{\tau}$, and $\tilde{S}_i$. We perform the 
inversion as specified in Eqs.~(57)--(62) of~\cite{mhd_code_paper}, but with a
slightly modified analytic quartic solver from the GNU Scientific
Library that outputs only the real roots.  We use the same technique as
in \cite{etienne08} to ensure that the values of $\tilde{S}_i$ and
$\tilde{\tau}$ yield physically valid primitive variables,
except we reset $\tilde{\tau}$ to
$10^{-10}\tilde{\tau}_{0,{\rm max}}$ (where
$\tilde{\tau}_{0,{\rm max}}$ is the maximum value of $\tilde{\tau}$
initially) when either $\tilde{S}_i$ or $\tilde{\tau}$ is 
unphysical (i.e., violate one of the inequalities~(34)~or~(35)
in \cite{etienne08}). The restrictions usually apply only to the region near
the puncture inside the horizon.

For each of our calculations, we set our outer boundary at $128 M$ and
use 8 AMR refinement levels.  The maximum resolution near each BH is
$\delta x/M = 0.03125$.  For our single BH test calculations, we place
our outer boundary at $128 M$ and use 6 AMR refinement levels.  For
these cases, the highest resolution near the BH is $\delta x/M = 0.0625$.

We model the emission of electromagnetic radiation by treating this
radiation loss as a perturbation, and neglect its influence on the
hydrodynamic flow, as well as any deviation from adiabaticity that it induces.  

\subsection{Diagnostics}
\label{sec:diagnostics}
\subsubsection{Surface density}
In order to track the global evolution of disk structure and compare with other disk calculations, it is useful
to define the surface density $\Sigma$.  Following \cite{shibata06}, we define
\begin{equation}
\Sigma(R,\phi)= \int_{z\ge0} \rho_0 u^t \sqrt{-g} dz \ ,
\end{equation}
where $R \equiv \sqrt{x^2+y^2}$ ($R$ will always be the cylindrical radius in this paper, while $r$ will always be the spherical polar radius).  We also report the angle-averaged surface density $\left< \Sigma(R)\right>$
where
\begin{equation}
  \left< \Sigma \right> \equiv \frac{1}{2 \pi} \int_0^{2\pi} \Sigma d
  \phi \ .
\end{equation}
\subsubsection{Flux diagnostics}
\label{sec:flux_diagnostics}
To derive meaningful flux diagnostics we must identify the conserved currents. Details of this derivation are given in
Appendix~A of \cite{farris10}.  To summarize, consider a 3D region $\Sigma_t$ which lies between
two world tubes $F$ and $L$ on a timeslice $t$=const.  Let $F$ be
defined by $h(t,x,y,z)=0$, and $L$ be defined by $l(t,x,y,z)=0$.  In
\cite{farris10}, this region is depicted as the lower shaded region
in Fig.~24.  For the purposes of this paper, we let $F$ be the world
tube defined by the apparent horizon(s), and $L$ be  the world tube
defined by a sphere of constant coordinate radius centered at the
origin. Of
course the surfaces could be chosen to take on other shapes as well.  
\subsubsection{Conserved quantities}
Now consider a conserved current, $j^{\mu}$ which satisfies
\begin{equation}
  \nabla_{\mu} j^{\mu} = 0 \ .
\end{equation}
Then it can be shown that (see e.g. Appendix A of \cite{farris10})
\begin{equation}
  \label{cons_eq}
  \dot{q} \equiv \frac{d q}{d t} = - \mathcal{F}_F + \mathcal{F}_L \ ,
\end{equation}
where
\begin{eqnarray}
q(t) &=& \int_{\Sigma_t} j^{\mu} d^3 \Pi_{\mu}\\
&=& \int_{\Sigma_t} j^{t} \sqrt{-g} \ d^3 x \ ,
\end{eqnarray}
\begin{eqnarray}
  \mathcal{F}_F &=& - \int_F \sqrt{-g'}j^h da db \ ,\\
  \mathcal{F}_L &=& - \int_L \sqrt{-g'}j^l da db \ .
\end{eqnarray}
Here $g'$ is the determinant of the metric in the $(t,h,a,b)$ or $(t,l,a,b)$
coordinate systems.  In the above example, $\mathcal{F}_F$ is the flux of
$q$ across the horizon(s), while $\mathcal{F}_L$ is the flux of $q$
across the outer sphere.  

\subsubsection{Freedom in coordinate choice}
These fluxes are independent of any changes in coordinates that leave
the slicing intact.  Equivalently, we may alter
the shift without affecting the flux, but the lapse must be kept the
same.  We can rewrite these fluxes in any other coordinate
system $(t,x,y,z)$ which preserves the same
slicing.  While $a$ and $b$ can be any two coordinates on the
surface, we label them here as $\theta$ and $\phi$ for convenience, as
this is done in our actual numerical calculations,
\begin{eqnarray}
  \mathcal{F}_F &=& - \int_F \sqrt{-g} \
  det\left|\frac{\partial(x,y,z)}{\partial(h,\theta,\phi)} \right|
  j^{\mu}\partial_{\mu}h \ d\theta d\phi \ ,\\
  \mathcal{F}_L &=& - \int_L \sqrt{-g} \
  det\left|\frac{\partial(x,y,z)}{\partial(l,\theta,\phi)} \right|
  j^{\mu}\partial_{\mu}l \ d\theta d\phi \ .
\end{eqnarray}
\subsubsection{Rest-mass conservation}
Rest-mass conservation, $\nabla_{\mu} (\rho_0 u^{\mu})=0$, corresponds to $j^{\mu}=\rho_0 u^{\mu}$. If we now define
\begin{eqnarray}
  M_0 &\equiv& \int_{\Sigma_t} \sqrt{-g} \rho_0 u^0 d^3 x =
  \int_{\Sigma_t}\rho_* d^3 x \ ,\\
  \mathcal{F}_F^{(M)} &\equiv& -\int_{F} \sqrt{-g} \ det\left|\frac{\partial(x,y,z)}{\partial(h,\theta,\phi)} \right|\rho_0 u^{\mu}h_{,\mu}  \
  d\theta d\phi  \ , \label{eq:FHM}\\
  \mathcal{F}_L^{(M)} &\equiv& -\int_{L} \sqrt{-g} \
  det\left|\frac{\partial(x,y,z)}{\partial(l,\theta,\phi)}
  \right|\rho_0 u^{\mu}l_{,\mu} \ d\theta d\phi \ , \label{eq:FLM}
\end{eqnarray}
then we may integrate Eq.~(\ref{cons_eq}) in time to derive the
following rest-mass conservation law
\begin{equation}
  \label{rest_mass_cons}
  \left(M_{0}(t) + \int_{t_i}^t dt \left( \mathcal{F}_F^{(M)} -
  \mathcal{F}_L^{(M)}\right)\right)/ M_{0,i}=1 \ ,
\end{equation}
where $M_{0,i}$ is the rest mass between the horizons and the surface
$L$ at $t=0$. Equivalently, we can define the rest-mass accretion rate
\begin{equation}
  \dot{M}_0 \equiv -\mathcal{F}_F^{(M)} +\mathcal{F}_L^{(M)} \ .
\end{equation}

\subsubsection{$E-\Omega J$ conservation}
We employ the helical Killing vector defined in Eq.~(\ref{killing_vec})
to construct another conserved current,
\begin{equation}
  \label{EmOmegaJ_current}
j^{\mu} \equiv \xi_{\nu} T^{\mu \nu}=T^{\mu}{}_t + \Omega
T^{\mu}{}_{\phi} \ .
\end{equation} 

We now define the following quantities
\begin{eqnarray}
  E(t) &\equiv& -\int_{\Sigma_t} \sqrt{-g} T^{t}{}_t  \ d^3 x \ ,\\
  J(t) &\equiv& \int_{\Sigma_t} \sqrt{-g}T^{t}{}_{\phi}  \ d^3 x \ ,
\end{eqnarray}

\begin{eqnarray}
  \mathcal{F}_F^{(E)} &\equiv& \int_{F} \sqrt{-g}\
  det\left|\frac{\partial(x,y,z)}{\partial(h,\theta,\phi)} \right|
  T^{\mu}{}_t  h_{,\mu}\ d\theta d\phi \ , \label{eq:FHE}\\
 \mathcal{F}_L^{(E)} &\equiv& \int_{L} \sqrt{-g}\
 det\left|\frac{\partial(x,y,z)}{\partial(l,\theta,\phi)} \right|
 T^{\mu}{}_t  l_{,\mu}\ d\theta d\phi \ , \label{eq:FLE}\\
 \mathcal{F}_F^{(J)} &\equiv& -\int_{F} \sqrt{-g} \
 det\left|\frac{\partial(x,y,z)}{\partial(h,\theta,\phi)} \right|
 T^{\mu}{}_{\phi} h_{,\mu}\  d\theta  d\phi \ ,\label{eq:FHJ}\\
 \mathcal{F}_L^{(J)} &\equiv& -\int_{L} \sqrt{-g} \
 det\left|\frac{\partial(x,y,z)}{\partial(l,\theta,\phi)} \right|
 T^{\mu}{}_{\phi} l_{,\mu}\  d\theta d\phi \ , \label{eq:FLJ}
\end{eqnarray}
and we see that Eqs.~(\ref{cons_eq}) and (\ref{EmOmegaJ_current}) give 
\begin{equation}
 \dot{E}-\Omega \dot{J} = -\left(\mathcal{F}_F^{(E)} - \Omega
  \mathcal{F}_F^{(J)} \right) + \left(\mathcal{F}_L^{(E)} - \Omega
  \mathcal{F}_L^{(J)} \right) \ ,
\end{equation}
for spacetimes possessing a helical Killing vector.  Again, we may integrate
in time to derive another conservation law

\begin{eqnarray}
  \label{E_minus_J_cons}
  E(t) - \Omega J(t) &&+ \int^t_{t_i} dt
  \left(\mathcal{F}_F^{(E)} - \Omega \mathcal{F}_F^{(J)} \right) \nonumber\\
&&- \int^t_{t_i} dt
    \left(\mathcal{F}_L^{(E)} - \Omega \mathcal{F}_L^{(J)} \right)\nonumber\\
    &=& E(t_i) - \Omega J(t_i)
\end{eqnarray}

\subsubsection{Spiral density wave diagnostics}
\label{sec:torque_density}
In our simulations, we use our CTS metric and Eq.~(\ref{gab_rotate}) to
ensure that our disk models exhibit quasistationary behavior before we
begin the binary inspiral.  Such quasistationary configurations are
interesting in their own right, as they lend insight into any accretion flow onto a binary before merger.  A key feature of this
flow is the presence of spiral density waves in the inner disk
cavity. Following \cite{macfadyen08}, we highlight the existence of
these density waves by calculating the surface density fluctuation
$\delta \Sigma$, defined by
\begin{equation}
  \delta \Sigma \equiv \frac{\Sigma -
    \left<\Sigma\right>}{\left<\Sigma\right>} \ .
\end{equation}
We also define the torque density, $dT / dR$, for comparison with
analytic models and other simulations,
\begin{equation}
  \label{torque_dens}
  \frac{dT}{dR} = \int \sqrt{-g} T^{\mu}{}_{\nu} \nabla_{\mu} \phi^{\nu} 
  R dz d \phi \ ,
\end{equation}
where $\phi^{\mu}\equiv (\partial_{\phi})^{\mu}$, which gives $\phi^{\mu}= (0,-y,x,0)$ in Cartesian coordinates. Details of the derivation of Eq.~(\ref{torque_dens}) are given in Appendix~\ref{torque_dens_appendix}.
\subsubsection{Luminosity diagnostics}
In order to study the electromagnetic emission from our disk
evolutions, we estimate the luminosity due to thermal bremmstrahlung
and nonthermal
synchrotron emission using the approximations described in
\cite{farris10}. For synchrotron emission, we assume the presence of a
small-scale, turbulent B field whose magnitude is approximated by
setting $P=\beta P_M \equiv \beta B^2/(8 \pi)$. We thus assume that
the magnetic pressure is some fraction $1/\beta$ of its equipartition
value. Simulations of magnetized accretion flows have demonstrated
that the magnetic fields do not typically reach their full
equipartition value \cite{mckinney04}. We have chosen $\beta=10$ to
account for this. We also assume that the radiation propagates through an
optically thin gas and we neglect the roles of radiation pressure and
radiative cooling on the hydrodynamic evolution.  While an accurate
estimation of the electromagnetic emission requires a full solution to
the radiative transfer problem, this crude method can provide a reasonable
estimate of the magnitude of the emission under suitable conditions. 

\subsection{Code tests}
\label{sec:code_tests}
Our HRSC general-relativistic hydrodynamic code has been 
thoroughly tested by passing a robust suite of tests. These tests include
maintaining stable rotating stars in stationary equilibrium, reproducing
the exact Oppenheimer-Snyder solution for collapse to a BH,
and reproducing analytic solutions for relativistic shocks and spherical Bondi accretion onto isolated BHs~\cite{mhd_code_paper}.  Our code has also been used to simulate the collapse of very
massive, rotating stars to black holes~\cite{liu07};
merging BHBH binaries~\cite{etienne07}, BHNS binaries \cite{etienne08,etienne09}, and relativistic hydrodynamic matter in the
presence of puncture black holes~\cite{faber07}.  Recently, our code has
been generalized to incorporate (optically thick) radiation transport
and its feedback on fluids in dynamical spacetimes~\cite{farris08}. 

Most of the above tests and simulations were performed 
on grids with uniform spacing. In some of the simulations, we 
utilized the multiple-transition fisheye transformation~\cite{campanelli06_fisheye} 
so that a uniform computational grid spacing 
corresponds to physical coordinates with spatially varying
resolution. Recently, we have modified our code 
so that we can use the moving-box AMR infrastructure provided by
Carpet~\cite{schnetter04}. To test our new code, we have performed 
shock-tube tests and 3+1 simulations of linear gravitational waves, single 
stationary and boosted puncture BHs, puncture BHBH binaries,  and rapidly and differentially rotating
relativistic stars.  Our AMR code has also been used to perform
simulations of BHNS mergers \cite{etienne09}, binary Bondi and binary
Bondi-Hoyle-Lyttleton accretion \cite{farris10} .
  
\begin{figure}
  \begin{center} 
    \epsfxsize=3.0in
    \leavevmode
    \epsffile{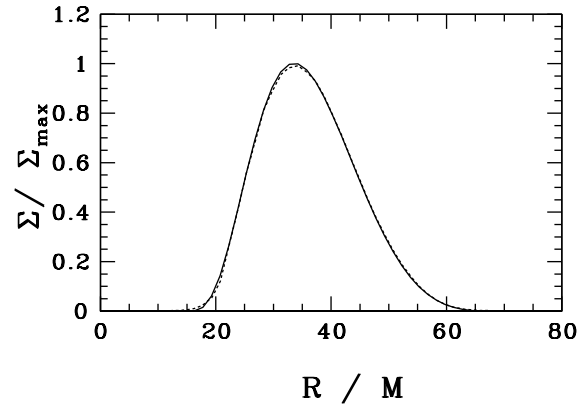}
  \end{center}
  \caption{Surface density profiles at $t=0$ (solid line) and
    $t\approx \ t_{disk}$ (dotted line), where $t_{disk}$ is the Keplerian period
    at the radius of maximum pressure.  Overlap indicates that disk
    accurately maintains equilibrium configuration over this time scale.}
  \label{fig:stable_equilib}
\end{figure}

All of our 3+1 AMR code tests were performed assuming equatorial
symmetry (i.e., symmetry about the $z=0$ orbital plane),  which we assume
in all evolutions presented in this paper.  We have checked that
our AMR code is able to accurately maintain a stable equilibrium disk
around a single BH, as demonstrated in
Fig.~\ref{fig:stable_equilib}.  For this test, we use the same disk
initial data as run A1 in Table~\ref{table:unmag}, except that we
set the background metric to be that of a single Schwarzchild BH at
the origin.  As we describe in Sec.~\ref{sec:initial_data}, such a
disk is an equilibrium solution and is expected to maintain its
initial profile.

\begin{figure}
\begin{center}
  \epsfxsize=3.0in
  \leavevmode
  \epsffile{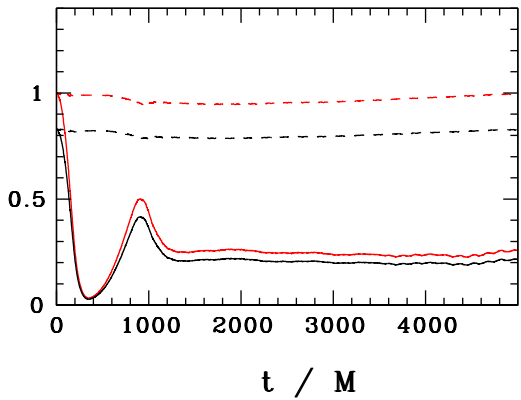}
\end{center}
\caption{Plots demonstrating the accurate maintenance of conserved
  quantities.  Rest mass $M_0(t) / M_{0,i} + \int dt \ (\mathcal{F}_F^{(M)} - \mathcal{F}_L^{(M)}  )/ M_{0,i}$
  (dashed red line) maintains its initial value accurately when
  compared to $M_0 / M_{0,i}$ (solid red).  Similarly, the conserved quantity
  $(E(t)-\Omega J(t) + \int dt \ (\mathcal{F}_F^{(E)} -
   \Omega \mathcal{F}_F^{(J)} ) 
 - \int dt \ (\mathcal{F}_L^{(E)} -
   \Omega \mathcal{F}_L^{(J)} ))/M_{0,i} $ (dashed black line)
   maintains its initial value accurately when compared to 
     $(E(t)-\Omega J(t))/M_{0,i}$ (solid black line).}
  \label{fig:flux_cons}
\end{figure}

We have also checked that the conservations laws in
Eqs.~(\ref{rest_mass_cons}) and (\ref{E_minus_J_cons}) are satisfied in a
quasistationary, binary spacetime, as described in Sec.~\ref{sec:diagnostics}.
In Fig.~\ref{fig:flux_cons}, the dashed red curve shows the left-hand side of
Eq.~(\ref{rest_mass_cons}), with the world tube $L$ chosen to
correspond to a sphere centered at the orgin with a radius $r_L=25 M$.
For comparison, we also plot $M_0(t)/M_{0,i}$ with the solid red
curve.  The data are from run A2, in which we impose a helical Killing
vector to solve for the metric as described in
Sec.~\ref{sec:early_epoch}, while evolving the hydrodynamics using
Eqs.~(\ref{hydro_evolution_rhostar}), (\ref{hydro_evolution_Stilde}),
and (\ref{hydro_evolution_tau}).   We find that Eq.~(\ref{rest_mass_cons}) is well satisfied,
indicating that our code is conserving rest mass correctly.
Similarly, the dashed black line shows the left-hand side of
Eq.~(\ref{E_minus_J_cons}), normalized by$M_{0,i}$, while the solid
black line shows $(E(t)-\Omega J(t))/M_{0,i}$ for comparison.  Again,
we see that our code is conserving $E-\Omega J$ correctly.

\section{Results}
\label{sec:results}
As discussed in Sec.~\ref{sec:basic_eqns}, we separate each of our
simulations into two phases.  We first perform early inspiral epoch
simulations in which we employ the quasistationary CTS metric while keeping
the BH separation constant.  This allows the disk to relax to a reliable
quasistationary state.  Upon achieving this state, we begin our
late inspiral and merger epochs simulations in which we evolve the metric in full GR, allowing
the BHs to inspiral and merge.  Parameters for each of these
disk runs are given in Table~\ref{table:unmag}.

\begin{figure*}
\vspace{-4mm}
 \begin{center}
    \epsfxsize=3.5in
    \leavevmode
  \epsffile{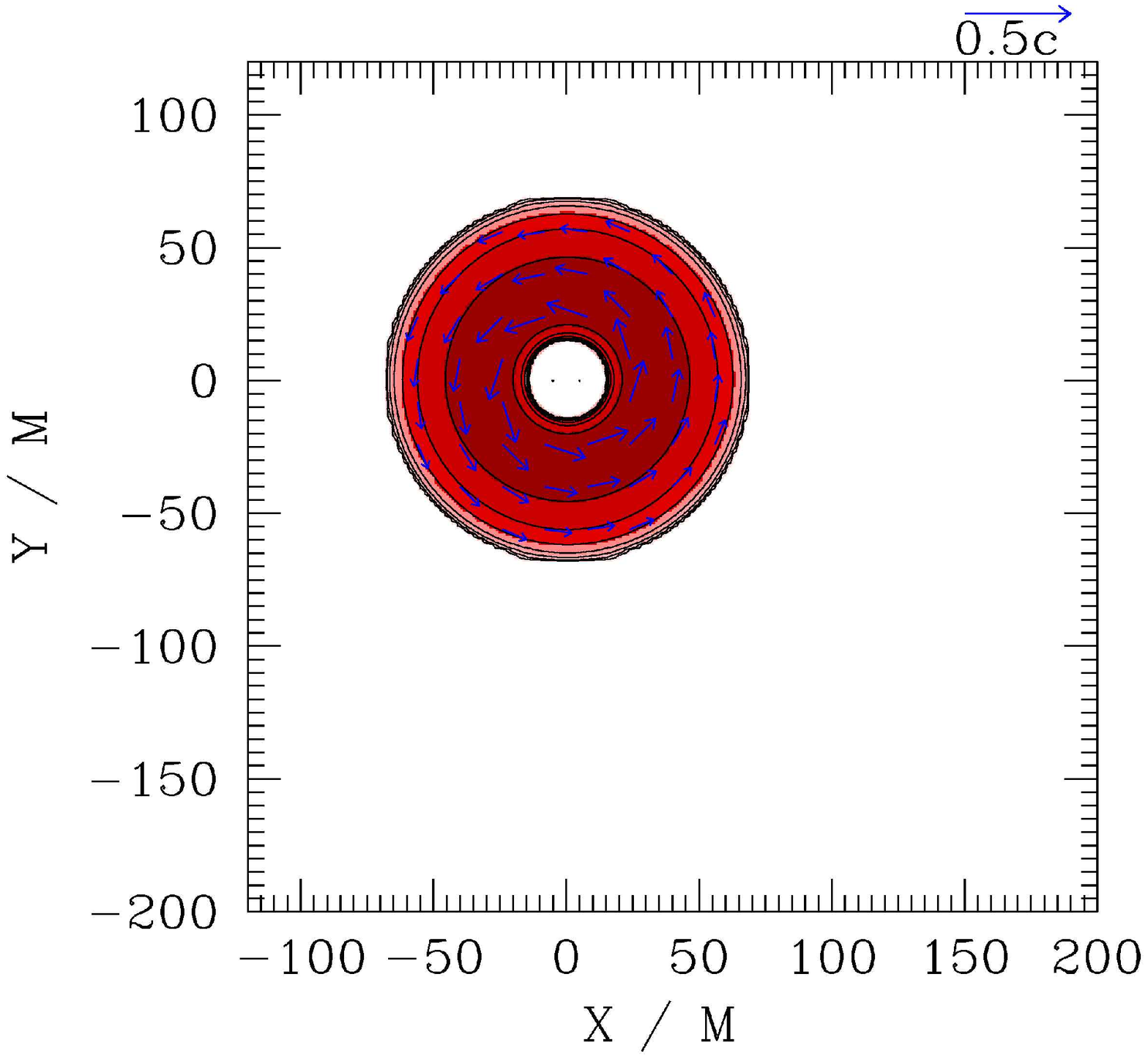}
\vspace{-4mm}
 \epsfxsize=3.5in
  \leavevmode
  \epsffile{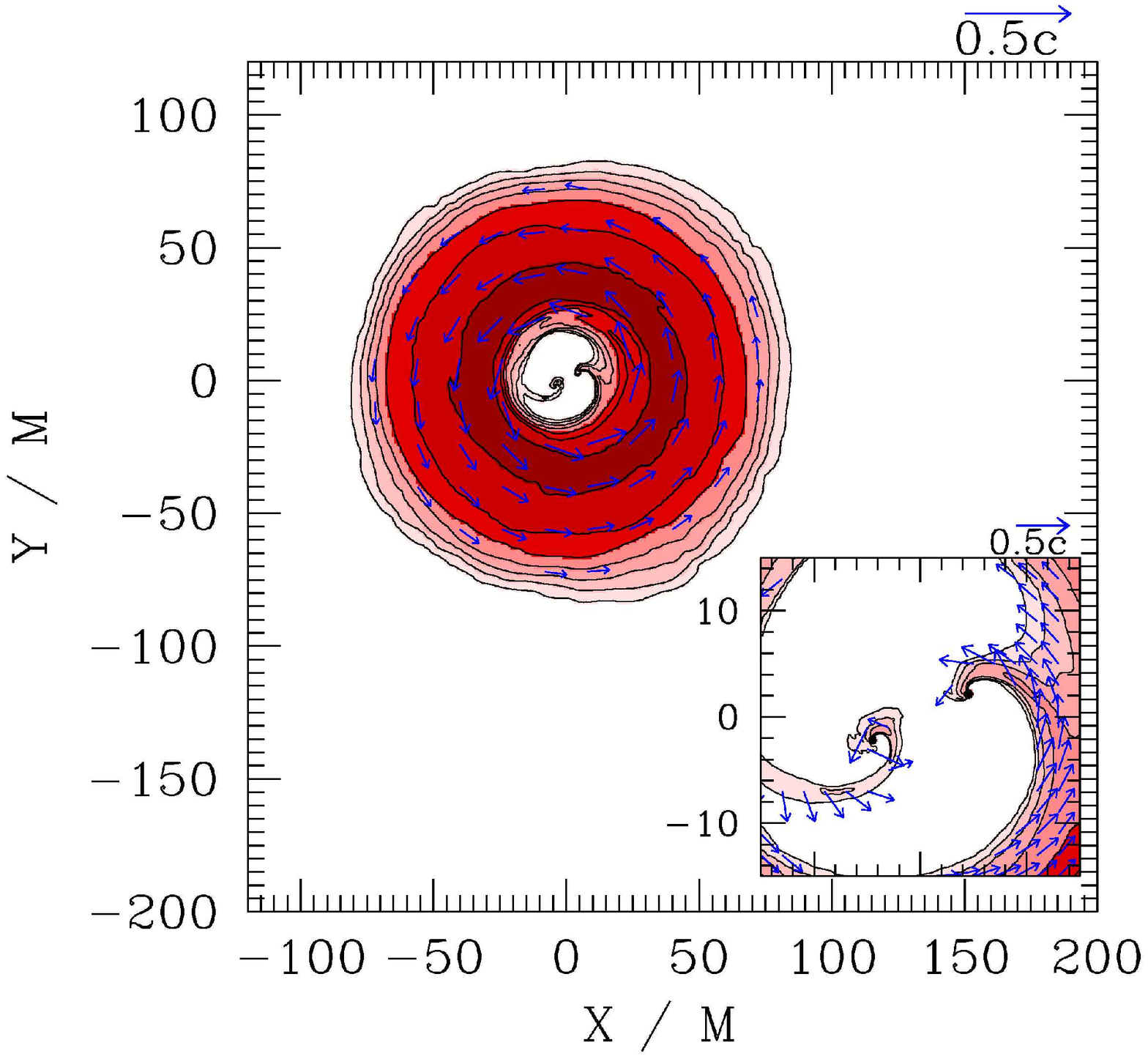}
  \epsfxsize=3.5in
  \leavevmode
  \epsffile{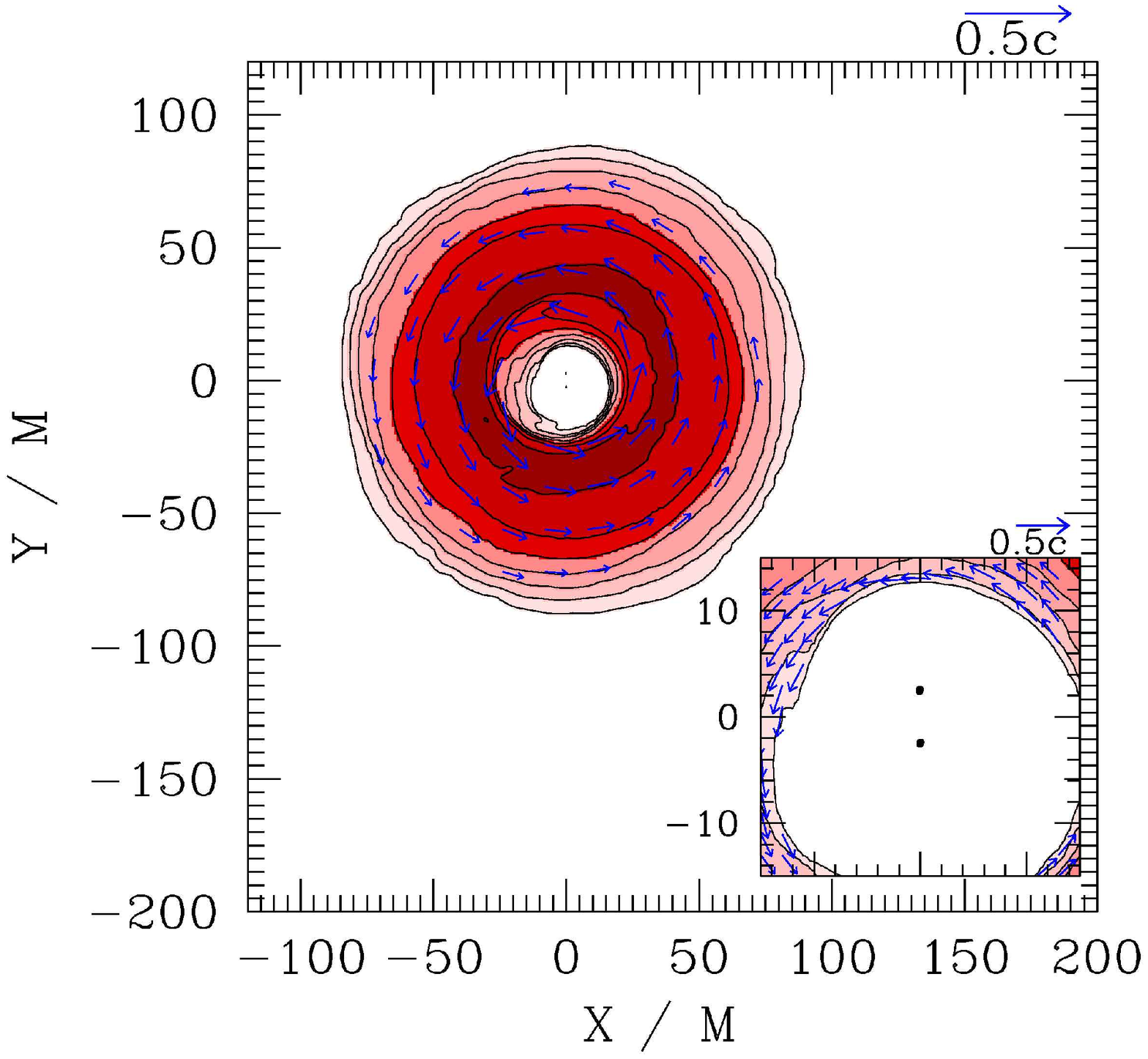}
  \epsfxsize=3.5in
  \leavevmode
  \epsffile{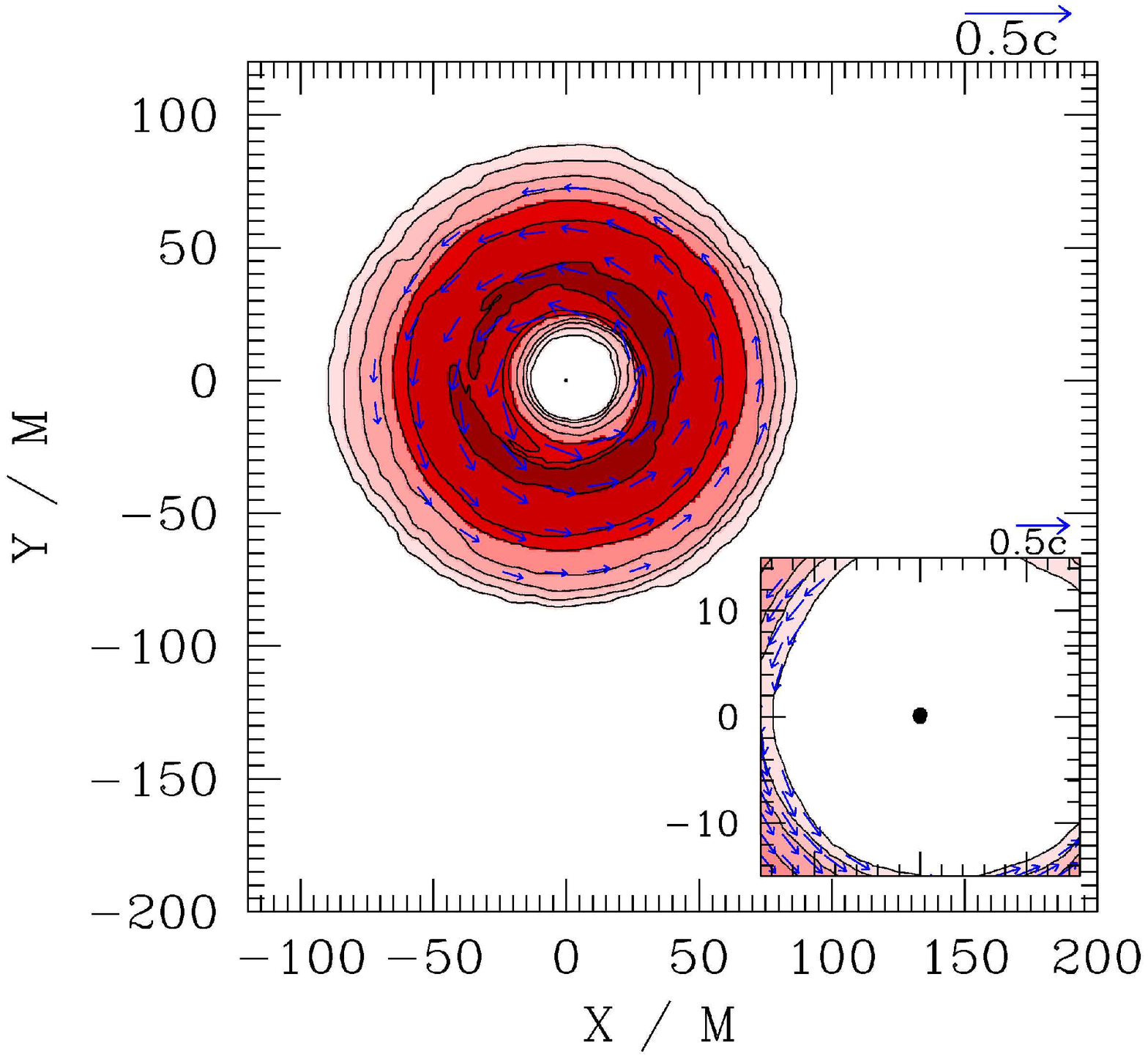}
\end{center}
\caption{Snapshots of rest-mass density $\rho_0$ contours in the
  orbital plane for cases with $\Gamma=5/3$.  Density contours are
  plotted at $\rho_0/\rho_{0,max}=10^{-2.75+0.5 j} \ \
  (j=1,2,....,6)$.  Contours of highest density are shown with darker
  shading and are near the BHs.  Blue arrows
  denote velocity vectors.  The apparent horizon interior is marked by
  a filled black circle.  The top left frame is the initial data from
  the early inspiral epoch, the top right frame is the relaxed,
  quasistationary disk, which serves as initial data for the late inspiral and
  merger epochs at $t \approx t_{merge}-1250M$.  The
  bottom left frame is the data from the late inspiral and merger epochs run at
  $t\sim t_{merge}-50M$, while the bottom left frame is the data from
  the late inspiral and merger epochs at
  $t\approx t_{merge}+t_{disk}$. }
\label{fig:dens_contour_5o3_xy}
\end{figure*}

\begin{figure}
\vspace{-50mm}
\begin{center}
  \epsfxsize=3.25in
  \leavevmode
  \epsffile{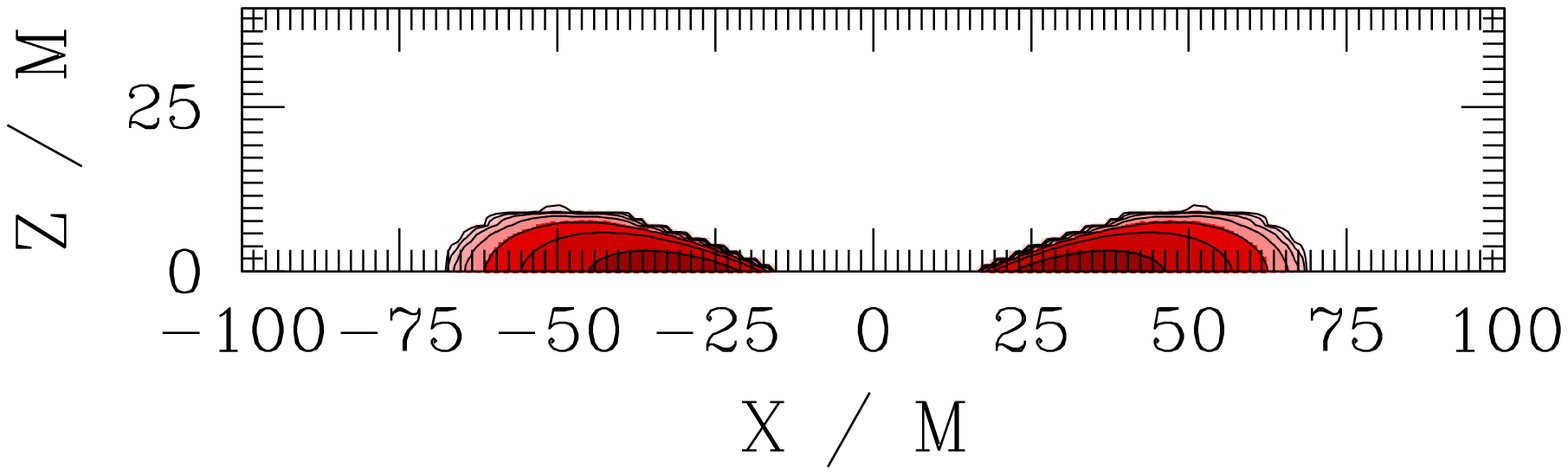}
\end{center}
\vspace{-50mm}
\begin{center}
  \epsfxsize=3.25in
  \leavevmode
  \epsffile{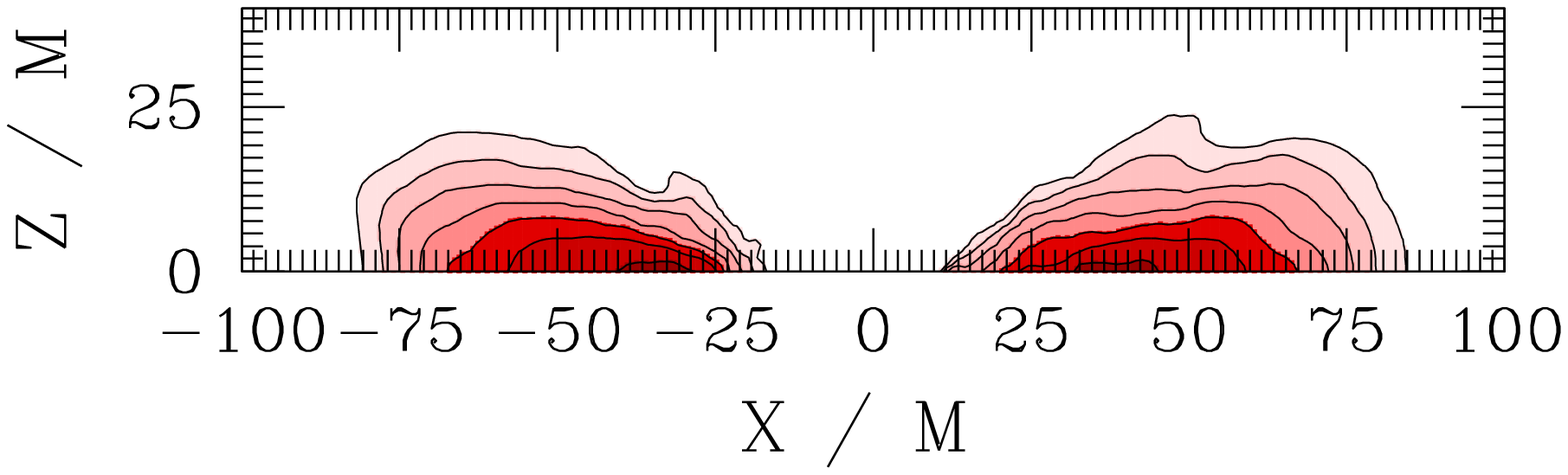}
\end{center}
\vspace{-50mm}
\begin{center}
  \epsfxsize=3.25in
  \leavevmode
  \epsffile{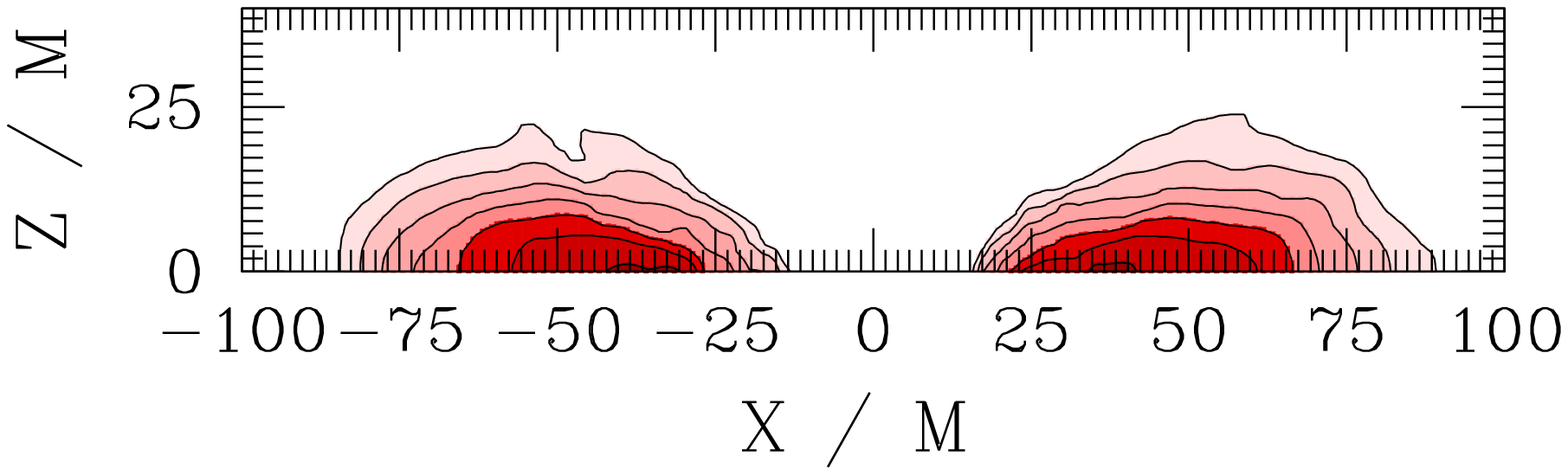}
\end{center}
\vspace{-50mm}
\begin{center}
  \epsfxsize=3.25in
  \leavevmode
  \epsffile{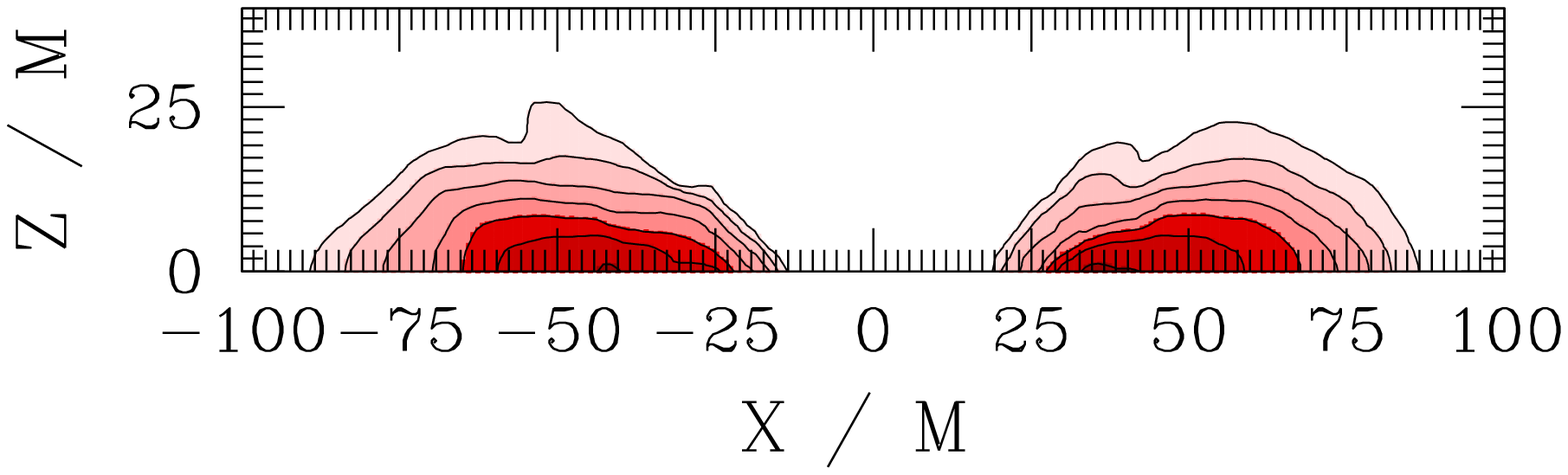}
\end{center}
\caption{Snapshots of rest-mass denstiy $\rho_0$ contours in the
  meridional plane for cases with $\Gamma=5/3$.  Density contours are
  plotted at $\rho_0/\rho_{0,max}=10^{-2.75+0.5 j} \ \
  (j=1,2,....,6)$.  Contours of highest density are near the BHs. The top frame is the initial data for
  the early inspiral epoch; the second frame is the relaxed,
  quasistationary disk that serves as initial data for the late inspiral and merger epochs;
  the third frame is the data from the late inspiral and merger epochs at
  $t\sim t_{merge}-50M$; the bottom frame is the data from the late inpiral
  and merger epochs at
  $t\approx t_{merge}+t_{disk}$. }
\label{fig:dens_contour_5o3_xz}
\end{figure}

Equatorial snapshots from our
simulations with $\Gamma=5/3$ can be seen in Fig.~\ref{fig:dens_contour_5o3_xy}, while meridional snapshots are
shown in Fig.~\ref{fig:dens_contour_5o3_xz}. The first two
snapshots are from the early inspiral epoch calculations, while the second
two snapshots are from the late inspiral and merger epochs calculations. We do not show snapshots for other equations of state here, as the accretion flow is qualitatively similar. Important results from simulations with other equations of state are reported in Table~\ref{table:lum_results}.

\begin{table}
\caption{Parameters for BHBH simulations}

\begin{tabular}{llllll}
Case \ \ \ &${}^{a}$Epoch \ \ \ \ \ \ \ \ \ \ \ &Orientation \ \ \ \ &
$\Gamma$ \ \ \ \ \  & ${}^{b} H/R$\\
\hline
\hline
A1&early inspiral&prograde&5/3&0.11\\
A2&&&4/3&0.08\\
A3&&&${}^{c}1.1$ &0.03\\
A4&&retrograde&4/3&0.08\\
\hline
B1&late inspiral&prograde&5/3&0.14\\
B2&and merger&&4/3&0.11\\
B3&&&${}^{c}1.1$&0.06\\
B4&&retrograde&4/3&0.11\\
\hline
\end{tabular}
\vskip 12pt
\begin{minipage}{12cm}
  \raggedright 
  ${}^{a}$ Initial binary separation $a/M=10$.

  ${}^{b}$ $H$ is the scale height of the disk at $R=\rdisk$

  (pressure max), measured at $t=0$ for case-A

  runs and at $t=t_{merge}$ for case-B runs.
 
 ${}^{c}$ Approximately isothermal.

\end{minipage}    
\label{table:unmag}
\end{table}

\subsection{Early inspiral epoch}
\label{sec:analytically_rotated}
While the formulation outlined in Sec.~\ref{sec:initial_data}
provides stable equilibrium disk initial data for a {\it single} BH,
torques from the {\it binary} disrupt this equilibrium.  Thus, it is
important to allow the gas to relax to a quasistationary state
before beginning the BH inspiral.  
We allow this relaxation to occur over $\sim 5 \tdisk$, where $\tdisk
\approx 1300 M$.  At this time, we find that
$\dot{M}_0$, $\Ls$, and $\Lb$ oscillate around roughly
constant values, and there is little evolution in surface density
profiles.  Here $\Ls$ and $\Lb$ refer to synchrotron and bremsstrahlung luminosity, respectively.

\begin{figure}
\begin{center}
  \epsfxsize=3.5in
  \leavevmode
  \epsffile{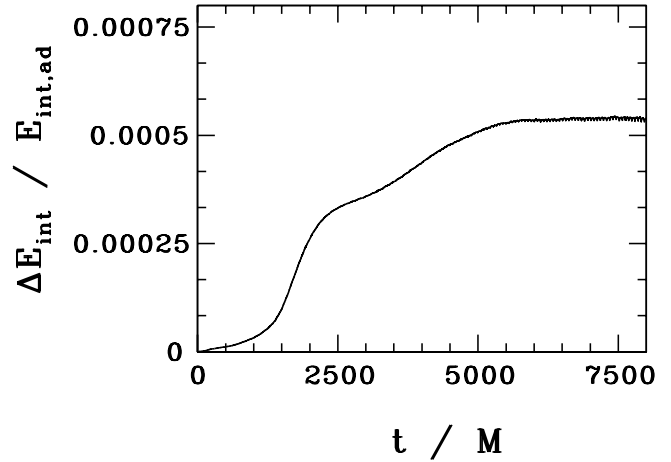}
\end{center}
\caption{Time evolution of $\Delta E_{int} / E_{int,ad}$ for
  run A2.}
\label{fig:internal_energy}
\end{figure}

During the relaxation process, the changes in the matter profiles
are due to the presence of binary torques.  These torques cause a
disruption of the inner edge of the disk, allowing some gas to fall onto
the BHs.  Infalling gas forms spiral waves and shocks which heat the gas near
the BHs.  
In the absence of shock heating, the disk would behave adiabatically, and
the internal energy density would be given by its polytropic value,
\begin{equation}
  \rho_0 \epsilon_{ad} = K \rho_0^{\Gamma} / (\Gamma-1) \ .
\end{equation}
Thus, shock heating may be
measured by computing the enhancement in the internal energy density above its
adiabatic value and integrating over the disk.  We therefore compute
\begin{equation}
  E_{int} = \int_{V_d} \sqrt{-g}\rho_0 u^t \epsilon \ d^3 x \ , 
\end{equation}
and
\begin{equation}
  E_{int,ad} = \int_{V_d} \sqrt{-g} \rho_0  u^t \epsilon_{ad} \ d^3 x \ .
\end{equation}
Here the integral is over $V_d$, which is the volume between $r= 10 M$
and the outer boundary of the computational domain at $128 M$.  This allows us to
compute $\Delta E_{int}$ in the bulk of the disk, ignoring the gas
near the BHs.  In Fig.~\ref{fig:internal_energy}, we plot $\Delta E_{int}/E_{int,ad}$ vs time
during the relaxation of the gas, where $\Delta E_{int} \equiv (E_{int}-E_{int,ad})$.  We find that $\Delta E_{int}/E_{int,ad}$
increases monotonically over the course of the relaxation process,
leveling off to a constant value of $\approx 5\times 10^{-4}$ as the disk reaches a
quasistationary state.  Because $\Delta E_{int} /E_{int,ad} \ll 1$, we conclude that shock heating does not
play a significant role in altering the bulk disk profile during this process.

\begin{figure}
\vspace{-4mm}
\begin{center}
  \epsfxsize=3.5in
  \leavevmode
  \epsffile{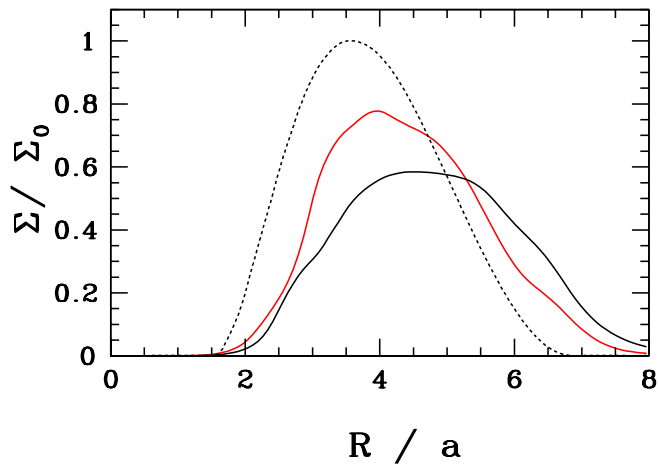}
\end{center}
\caption{Surface density profiles of the $\Gamma=5/3$ disk as a function of radius. The
  dashed black curve is the initial disk density
  proﬁle; the solid red one is the surface density proﬁle when the disk
  has reached a quasistationary configuration after $t\gtrsim 5 \tdisk$,
  averaged over $\sim 2 \torb$.  The solid black curve is the density profile
  following the merger, averaged over $\sim 2 \torb$. $\Sigma_0$ is the initial maximum surface density.}
\label{fig:surf_dens_5o3_plot}
\end{figure}

In order to get a sense of the change in the disk profile from the
initial data, we plot angle-averaged surface density profiles (see
Fig.~\ref{fig:surf_dens_5o3_plot}). In each case, the torque from the
binary has the overall effect of pushing matter outward.   
However, the
effect of the torques diminishes once the disk matter has
moved outward and a quasistationary state is achieved after $t\sim 5 \tdisk$.

\begin{figure}
\vspace{-4mm}
\begin{center}
  \epsfxsize=3.5in
  \leavevmode
  \epsffile{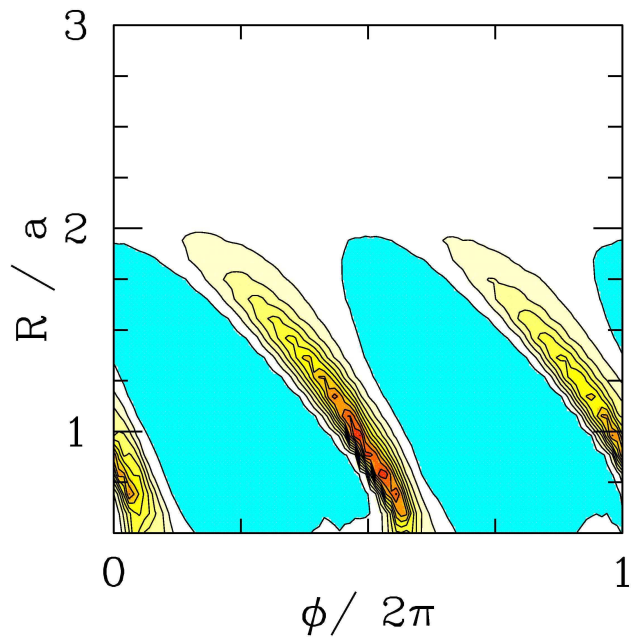}
\end{center}
\caption{Time-averaged surface density fluctuation 
\break
$(\Sigma-\left<\Sigma\right>)/\left<\Sigma\right>$ in
  the rotating frame in which the binary is at rest. The binary point masses are located at $R/a=0.5$ and $\phi=(0,\pi)$.
  Red regions are density maxima and blue regions are density minima.
  Data is from run A2 with $\Gamma=4/3$.}
\label{fig:surf_dens_4o3_polar_plot}
\end{figure}

\begin{figure}
  \vspace{-4mm}
  \begin{center}
    \epsfxsize=3.5in
    \leavevmode
    \epsffile{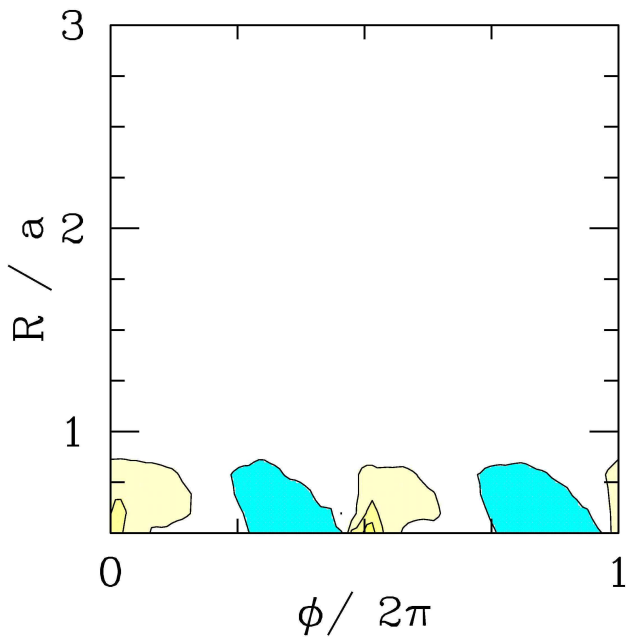}
  \end{center}
  \caption{Same as Fig.~\ref{fig:surf_dens_4o3_polar_plot} but for a
    retrograde disk.}
  \label{fig:surf_dens_retro_polar_plot}
\end{figure}

In Fig.~\ref{fig:dens_contour_5o3_xy}, we plot snapshots of density
contours for case A1. We see that the disk cavity, which initially extends
to $R=15M$, becomes partially filled, with a clear spiral structure, at
small radii $R \lesssim 2 a = 20 M$ for
each of the prograde cases. A similar spiral arm structure extending inside the disk cavity has been seen in Newtonian simulations (cf. \cite{macfadyen08,hayasaki07,roedig11}). In the retrograde case (A4), this
structure is largely absent.  This point is further emphasized by
comparing the surface density fluctuation $\Delta \Sigma \equiv
(\Sigma - \left<\Sigma\right>) / \left<\Sigma\right>$ between the
prograde case A2 and the retrograde case A4 (see Figs.~\ref{fig:surf_dens_4o3_polar_plot} and \ref{fig:surf_dens_retro_polar_plot}).  Following
\cite{macfadyen08}, we compute this quantity in a rotating frame in
which the binary is stationary, and average over several $\torb$.  In
Fig.~\ref{fig:surf_dens_4o3_polar_plot}, we clearly see two strong
spiral arms emanating from each BH and extending throughout the
cavity region.  However, unlike the results of the 2D, thin-disk simulations
presented in \cite{macfadyen08}, we do not see spiral density
waves extending into the bulk of the disk.  We attribute this to the fact
that our 3d disks are thicker, which allows waves that are initially propagating in the
radial direction to be deflected in the vertical direction, disrupting
the spiral density wave structure.  Such effects have been
demonstrated even for geometrically thin disks in which density and
temperature are stratified in the z direction \cite{lin90a,lin90b}.

\begin{figure}
\vspace{-4mm}
\begin{center}
  \epsfxsize=3.5in
  \leavevmode
  \epsffile{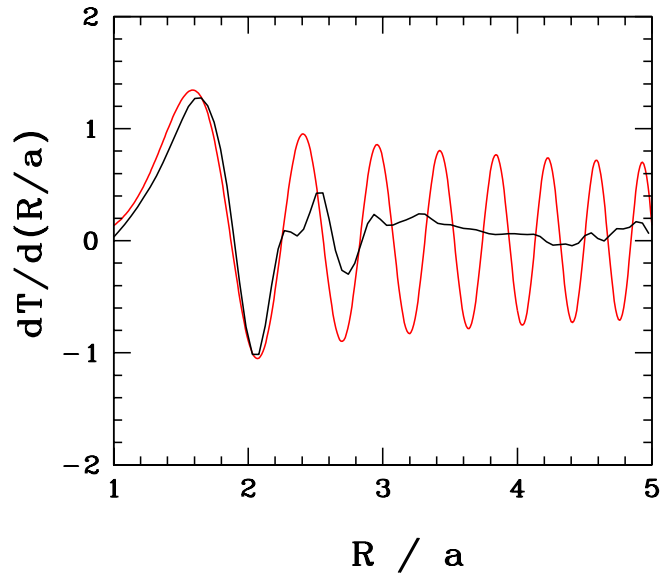}
\end{center}
\caption{
Time-averaged torque density $dT/dR$ exerted by the binary on the disk
after $t \gtrsim 5 \tdisk$. Time averaging was
carried out over $\sim 2 \torb$ after the disk has reached a
quasistationary state.  Data from run A2 with $\Gamma=4/3$. The torque is plotted in units of $10^{-3} M a \Sigma_0$.}
  \label{fig:torque_dens_4o3_plot}
\end{figure}

The structure of the spiral density waves observed near $\rin$ is roughly consistent with
the theoretical Newtonian picture of a wave that is excited by the
binary torques at the
outermost Lindblad resonance at $R_2 = (3/2)^{2/3}a$ with orbital resonance $\Omega_{bin}:\Omega=3:2$ (initially, $R_2 = 0.87 \rin$). We
demonstrate this by computing the angle-averaged torque density as
described in Sec.~\ref{sec:torque_density}, and comparing with the (Newtonian)
analytic prediction given in Eq.~(31) of \cite{macfadyen08},
\begin{equation}
  \frac{dT}{dR} \approx \frac{49}{288} \pi^2 \Sigma(R_2)
  \Omega_{bin}^2 \frac{a^4}{\lambda_2}
  \mbox{Ai}\left(\frac{R_2-R}{\lambda_2}\right) \ ,
\end{equation}
where $\lambda_2 = 2^{-2/3}(H/R)^{2/3}a$.  Based on numerical data
from run A1 at $t\gtrsim 5 \tdisk$, we choose $\Sigma(R_2)=4 \times
10^{-4} \Sigma_0$ and
$H/R=0.3$, where $\Sigma_0$ is the initial maximum surface density and
$H/R$ is measured at $R_2$. As we show in Fig.~\ref{fig:torque_dens_4o3_plot}, we find very good
agreement with the analytic prediction out to $R/a \approx 2.2$, but
break down at larger radii.  This breakdown is not unexpected, as we
have argued above that spiral density waves do not extend into the
bulk of the disk as a result of the thickness of our 3d disks.

\begin{figure}
\vspace{-4mm}
\begin{center}
  \epsfxsize=3.5in
  \leavevmode
  \epsffile{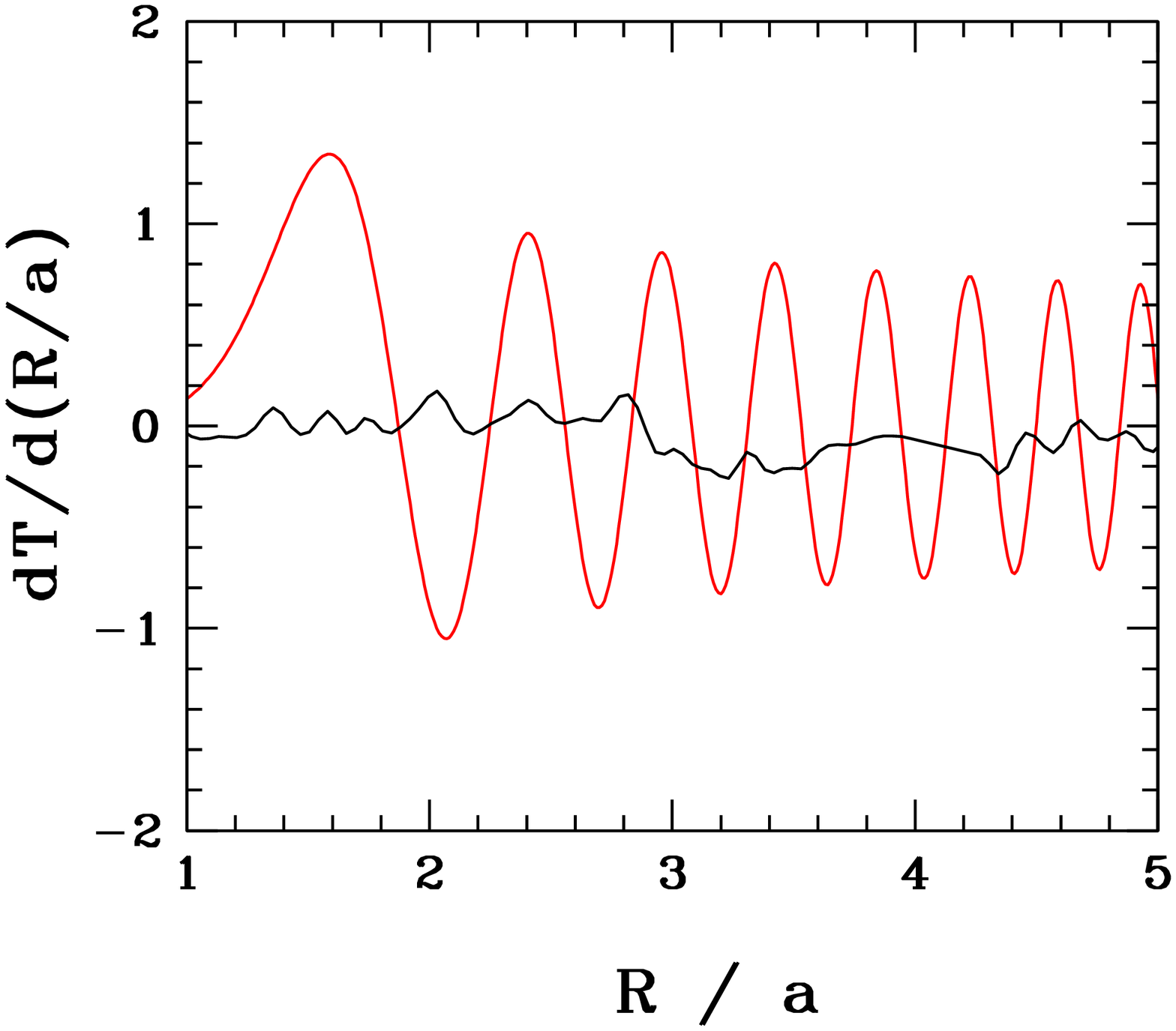}
\end{center}
\caption{Same as Fig.~\ref{fig:torque_dens_4o3_plot}, but for
  retrograde disk (run A4).}
  \label{fig:torque_dens_retro_plot}
\end{figure}

For the retrograde case (A4), we find that the spiral density waves
are largely absent, as shown in
Figs.~\ref{fig:surf_dens_retro_polar_plot} and
\ref{fig:torque_dens_retro_plot}.  This is expected, as the
Lindblad resonance does not exist when the angular momentum of the
disk and the binary are antialigned.

\subsection{Late inspiral and merger epochs}
Having allowed the disk to relax to a quasistationary state, we
turn to the fully relativistic evolution of metric and matter fields
in order to investigate variations in electromagnetic
luminosity over the course of the inspiral and merger.  Our
calculations in this epoch apply to the decoupling phase of binary
inspiral, through merger, but before appreciable gas fills the hollow
due to viscosity. We again
consider the prograde cases with $\Gamma=4/3$ (B1), $\Gamma=5/3$ (B2), and $\Gamma=1.1$ (B3)
the retrograde case with $\Gamma=4/3$ (B4).  In each case, we use the
relaxed data from the end of the corresponding quasistationary
metric run as initial data.

\begin{figure}
  \begin{center}
    \epsfxsize=3.0in
    \leavevmode
    \epsffile{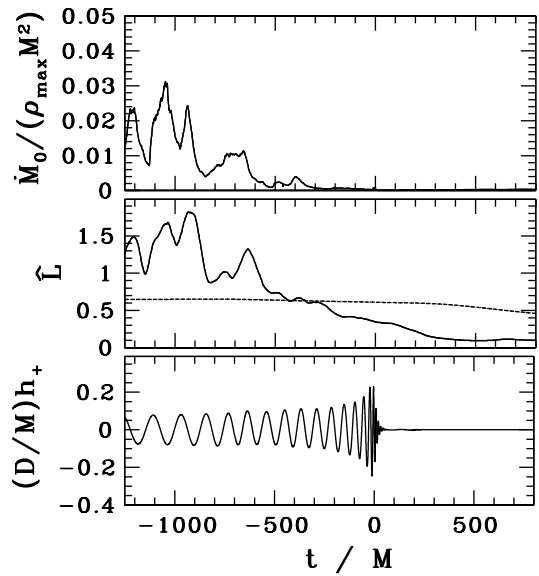}
  \end{center}
  \caption{Time evolution of total BH accretion rate across the BH horizons $\dot{M}_0$,
    luminosity $\hat{L}$ and waveform
    $D h_{+}$ for a circumbinary prograde disk with $\Gamma=5/3$.  The initial binary separation is $a= 10M$
    and the BHs evolve to merger. $\dot{M}_0/(\rho_{max} M^2)$ is the
    dimensionless accretion rate. Here, $\rho_{max}M^2=0.2 n_{12}M_8^2 M_{\odot} \mbox{yr}^{-1}$.  $\hat{L}\equiv L /[10^{46} M_{8}^3 n_{12}^2 \mbox{ erg
      s}^{-1}]$ is the total luminosity due to
    bremsstrahlung (dashed line) and synchrotron (solid line)
    emission. For synchrotron emission, we assume $\beta=10$. $h_{+}$ is the $+$ polarization of the gravitational
    wave signal as measured by an observer looking down the polar axis
    at a distance $D$ from the binary.  BHBH merger occurs at $t=0$.}
  \label{fig:time_evol_5o3}
\end{figure}

As the binary inspiral proceeds and the separation shrinks, the
torques due to the binary are diminished.  As a result, we find that
the spiral density waves visible when $a/M=10$ have largely
disappeared by the time of merger and remain absent after the merger.  This is evident in the bottom left
and right frames of Fig.~\ref{fig:dens_contour_5o3_xy},
which show snapshots of the density in the equatorial plane $\sim 50M$
before merger, and $\sim 1 \tdisk$ after
merger.  This effect is also illustrated by the evolution of
$\dot{M}_0$, $\Lb$, and $\Ls$ in
Fig.\ref{fig:time_evol_5o3}. Here, we have computed the luminosity assuming a fiducial value of $n_{disk}=10^{12} \mbox{ cm}^{-3}$, where $n_{disk}$ is the baryon number density at $\rdisk$.  This value is consistent with density estimates for a typical AGN derived from the Shakura-Sunyaev disk model \cite{shakura73,novikov73,shapiro_book_83}, albeit in a radiation-dominated, geometrically thin regime. However, because there are large variations in the gas densities in galactic cores, we provide density scalings for our results.  Because the position of the $m=2$ outermost Lindblad resonance
is approximately given by $R_2 \approx (3/2)^{2/3}a$, we see that as
the binary separation is reduced, the location of the resonance
retreats farther inside $\rin$, enabling less
matter to be stripped from the inner edge of the disk, and reducing
$\dot{M}_0$.  The reduction in accretion similarly suppresses the
electromagnetic luminosity generated near the BHs.  This effect is
exacerbated by the reduction in shock heating due to binary torques,
as this lowers the temperature of the gas and reduces emissivities.
Each of these effects is reflected in
Fig.~\ref{fig:time_evol_5o3}.  We also show the
$h_{+}$ polarization amplitude of the accompanying gravitational wave for
comparison.  Evidently, the decrease in electromagnetic luminosity
beginning at the onset of decoupling is a precursor to the late
inpiral and merger gravitational radiation. We find that the choice of EOS
can play a significant role in setting the magnitude of the accretion
rate and the luminosity.  Larger values of $\Gamma$ lead
to both larger $\dot{M}_0$ as well as higher luminosities.  We tabulate the values of luminosities, accretion rates
and characteristic frequencies of emission at the
onset of the late inspiral and merger epochs and just prior to  merger in
Table~\ref{table:lum_results}.   
In addition to the increase in the amount of gas near the BHs, larger values of
$\Gamma$ also allow the gas to be shock heated more effectively.  Because
the bremsstrahlung and synchrotron emissivities are sensitive to
temperature, this also leads to an increase in luminosity.
Comparing Eq.~(\ref{synch_emissivity}) and Eq.~(\ref{brem_emissivity})
below, we
see that the temperature dependence is much stronger for synchrotron
emission.  This can explain the particularly large differences in synchrotron
luminosity for the different cases reported in
Table~\ref{table:lum_results}.  This effect also leads the synchrotron
luminosity to be dominated by emission from the heated region near the
binary, whereas the 
bremsstrahlung emission is predominantly from the bulk of the
disk. This dependence accounts for the high variability of the synchrotron
luminosity in comparison to that of the bremsstrahlung emission. 

Because our simulations assume a perfect fluid with no dynamical magnetic
fields (turbulent fields are assumed only to estimate synchrotron emission), there is no viscosity present to counteract the effect of the
binary torques in driving matter outward.  As a result, we find that
even after relaxing to a quasistationary disk state in our early inspiral
epoch calculations, in which the accretion rate and luminosity
oscillate around fixed values, there remains an overall slow outward
drift in the bulk of the disk.  This is evident in
Fig.~\ref{fig:surf_dens_5o3_plot}, in which the solid red curve shows
the surface density profile at the beginning of the binary inspiral,
while the solid black curve shows the surface density profile at $t \sim
\tdisk$ after the merger.  We see clear evidence that the bulk of the
disk moves outward, although we
suspect that this effect may be altered by the inclusion of viscosity.

\begin{figure}
\begin{center}
  \epsfxsize=2.75in
  \leavevmode
  \epsffile{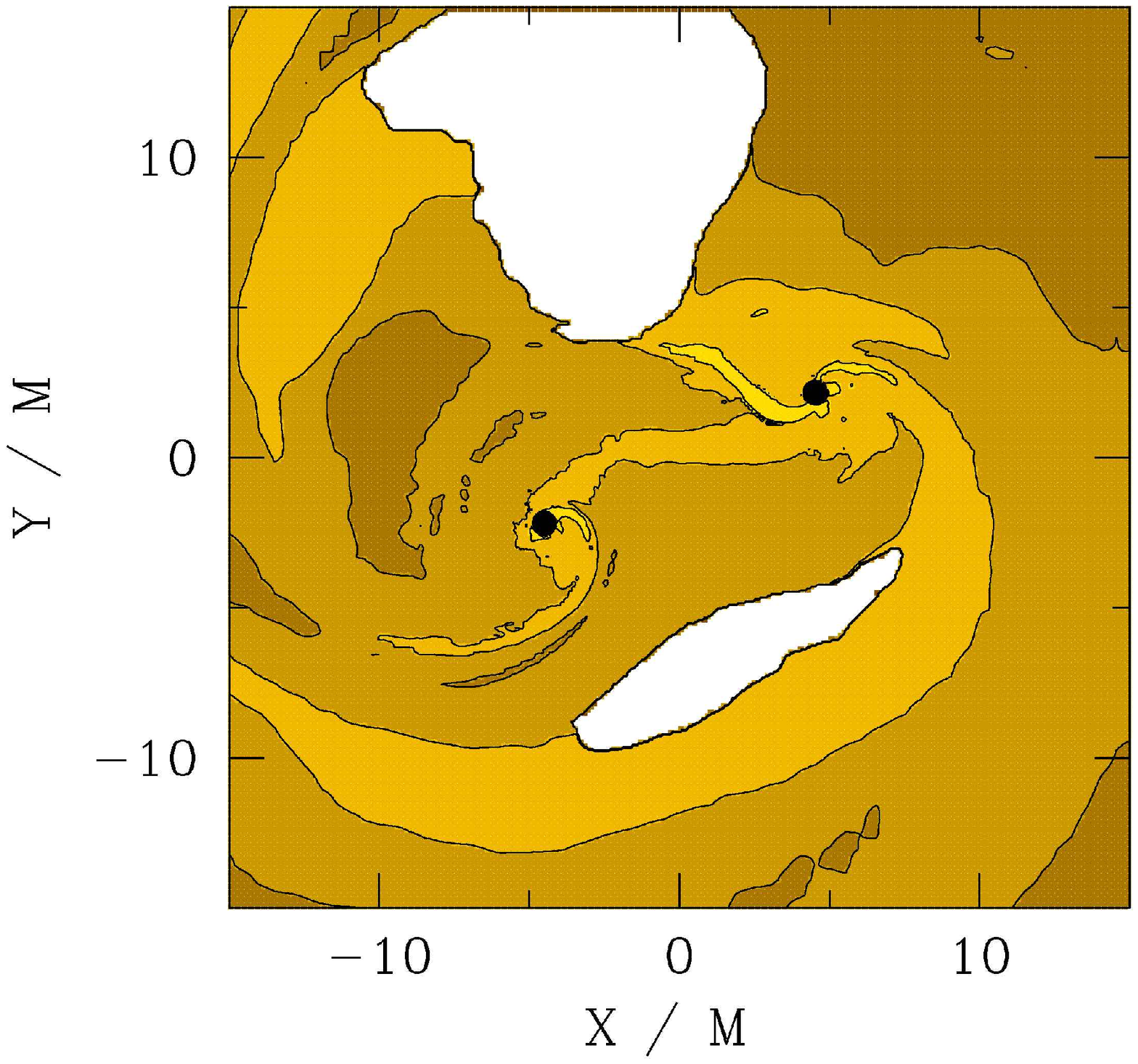}
\end{center}
\begin{center}
  \epsfxsize=2.75in
  \leavevmode
  \epsffile{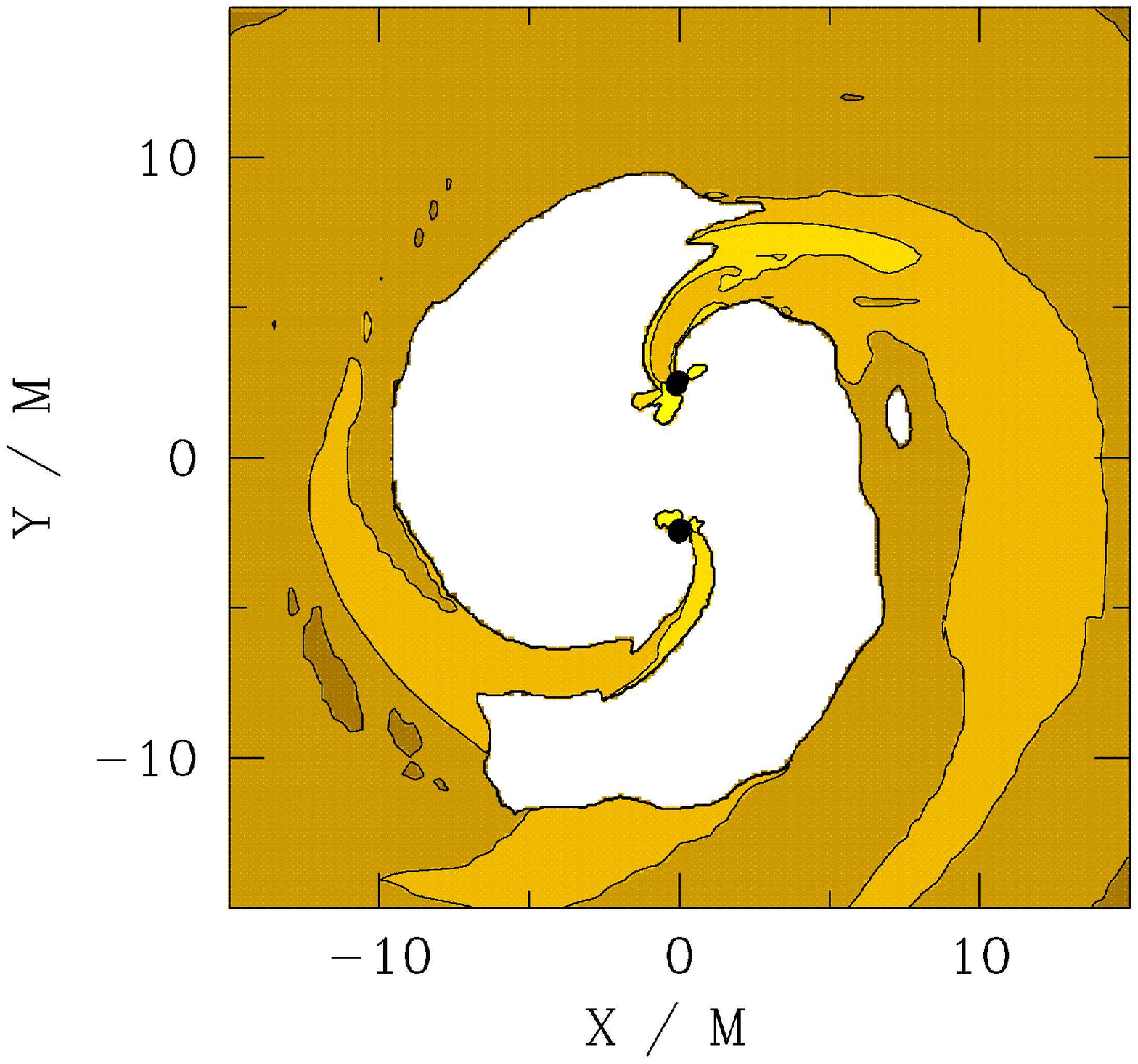}
\end{center}
\begin{center}
  \epsfxsize=2.75in
  \leavevmode
  \epsffile{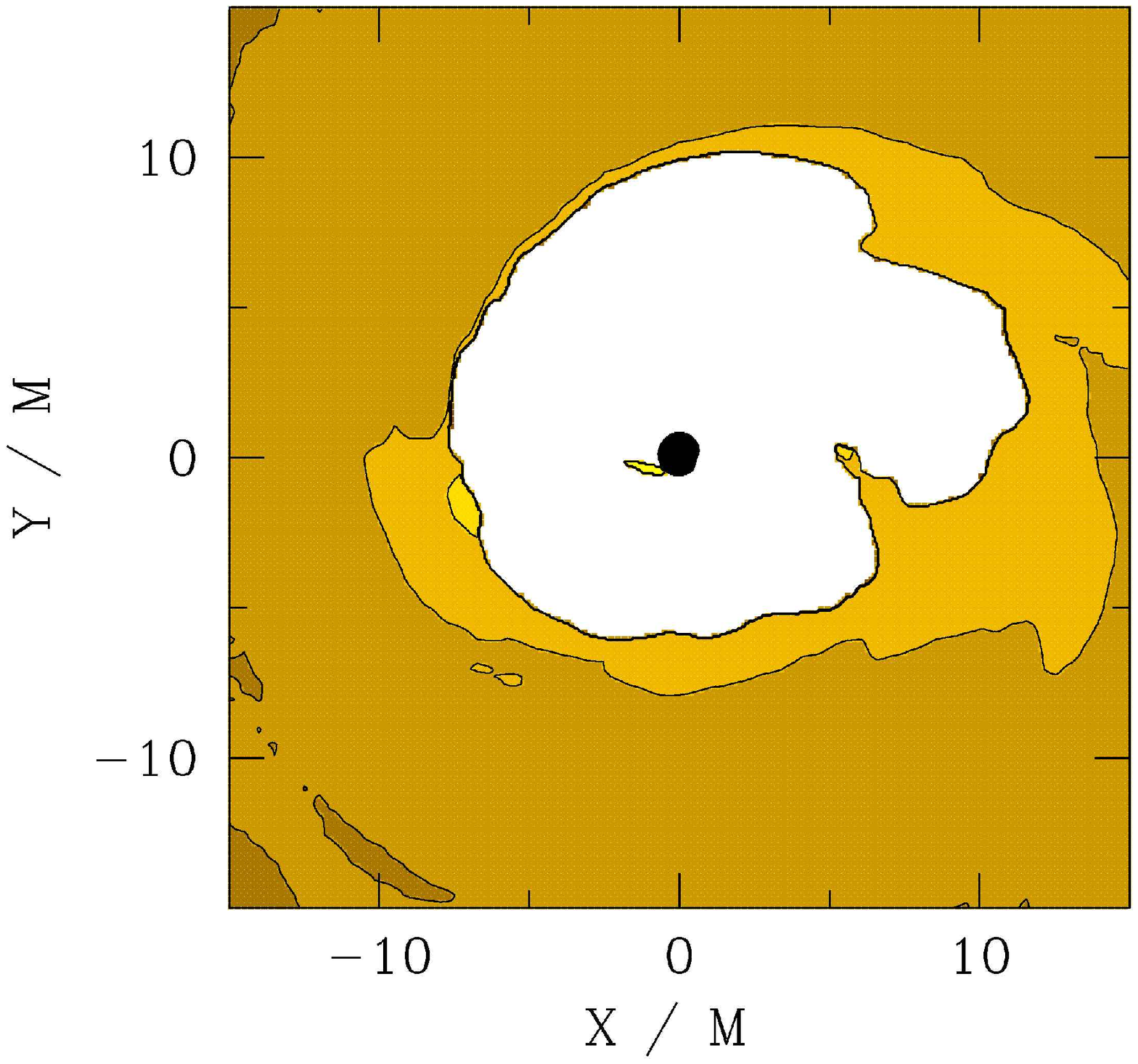}
\end{center}
\caption{Contours of temperature $kT$ (=  $ m_B P / \rho_0$) at select
  times for prograde disk with $\Gamma=5/3$.
  Contours correspond to $kT=940 \times 10^{-5+0.33 j} MeV\ \  (j=1,2,..
..,6)$. Frames correspond to the beginning of the late inspiral and
merger epochs
phase at $t\approx t_{merge}-1250M$ (top),
$t\sim t_{merge}-50M$ (middle), and $t\sim t_{merge}+\tdisk$ (bottom). Regions with
density less than $\rho_0/\rho_{0,max} < 10^{-4.5}$ are left
white. Lighter shading denotes higher $kT$.}
  \label{fig:temp_contour_5o3}
\end{figure}

In Paper I, we demonstrated that shocks near the BH horizons increased
in strength throughout the merger as the BHs move more supersonically
through the surrounding gas. This shock strengthening leads to a temperature increase in the
inner region, which in turn gives rise to an increasing luminosity
peaking at the moment of merger.  Such a temperature increase is
largely absent in the disklike accretion case treated here, as can be seen by
comparing temperature contours at the beginning of the late inspiral
and merger epochs
simulation and at $t\sim t_{merge} - 50 M$, as displayed in
Fig.~\ref{fig:temp_contour_5o3}.  As such, we do not expect to see significant
increase in luminosity during the post-decoupling inspiral phase, even at high-frequency components of the spectrum.  We note that this contrasts the
conclusions of the Newtonian calculation in \cite{chang10}, where a
brightening of the precursor light curve before merger is found.  However, these
results do not necessarily contradict one another, as \cite{chang10} consider geometrically
thin, optically thick disks around non-equal-mass BHBH binaries. This
is a very different scenario from the geometrically thick, optically thin disks
surrounding equal-mass binaries that we consider in this paper.

\begin{figure}
\begin{center}
  \epsfxsize=2.75in
  \leavevmode
  \epsffile{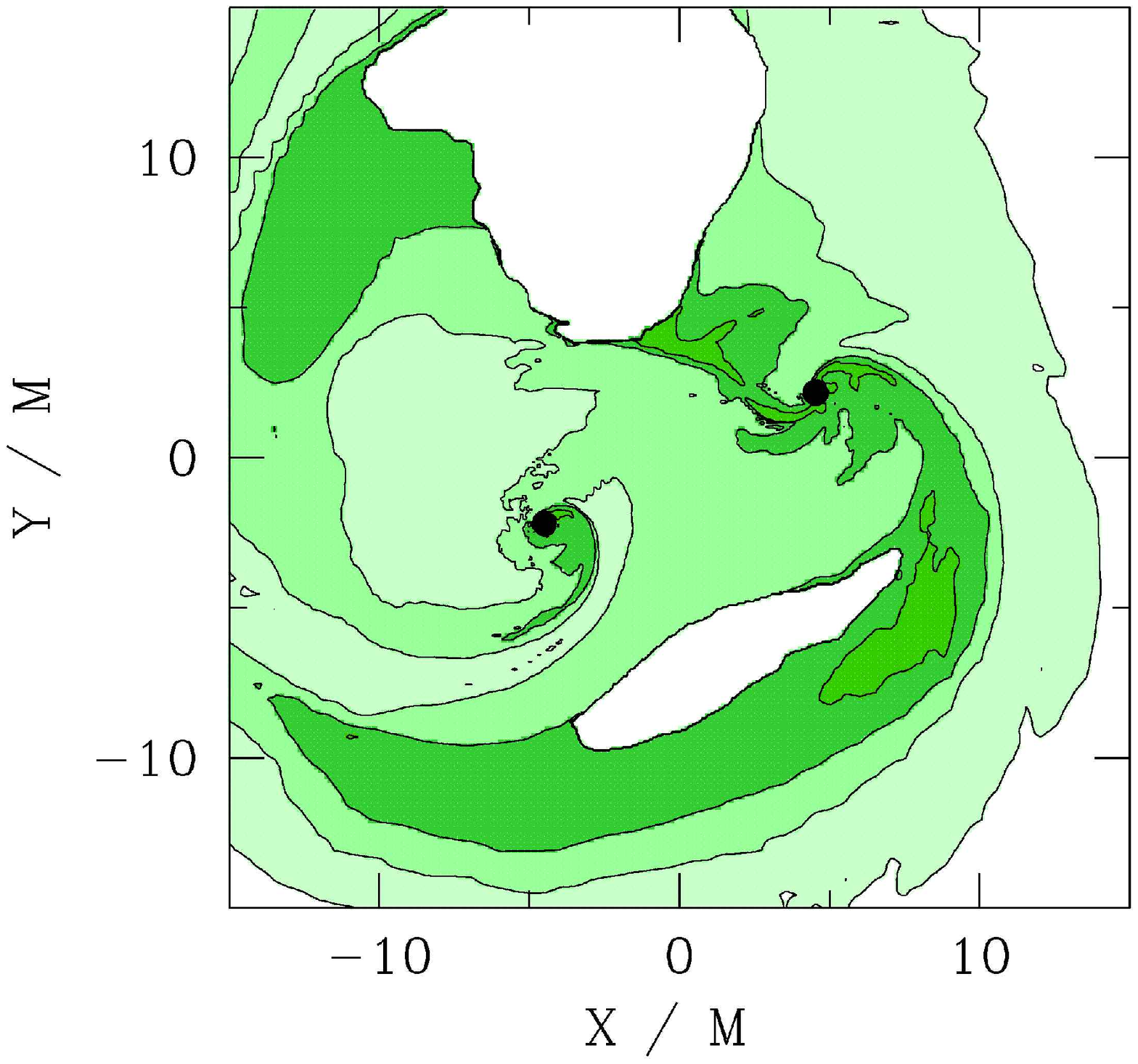}
\end{center}
\begin{center}
  \epsfxsize=2.75in
  \leavevmode
  \epsffile{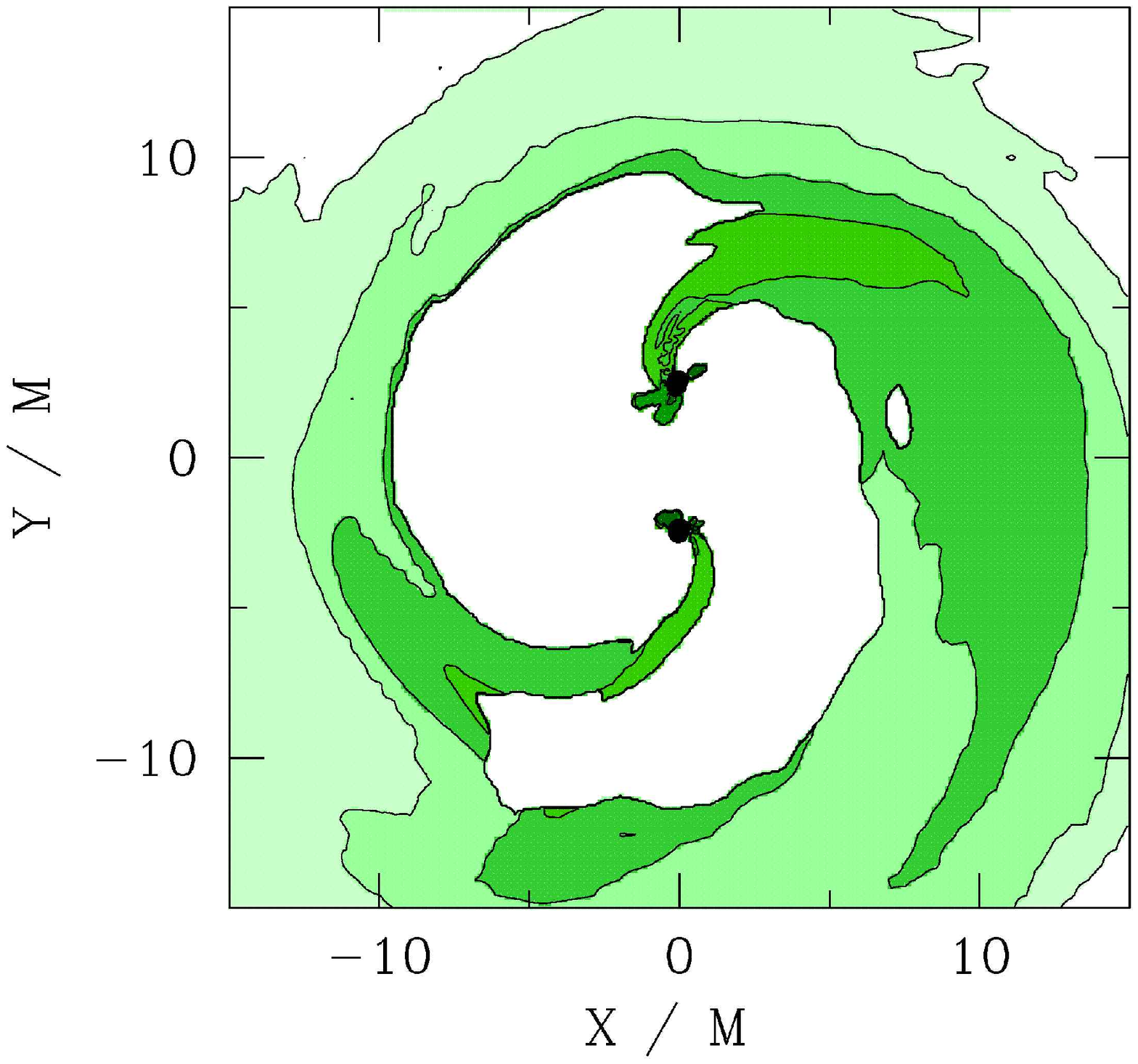}
\end{center}
\begin{center}
  \epsfxsize=2.75in
  \leavevmode
  \epsffile{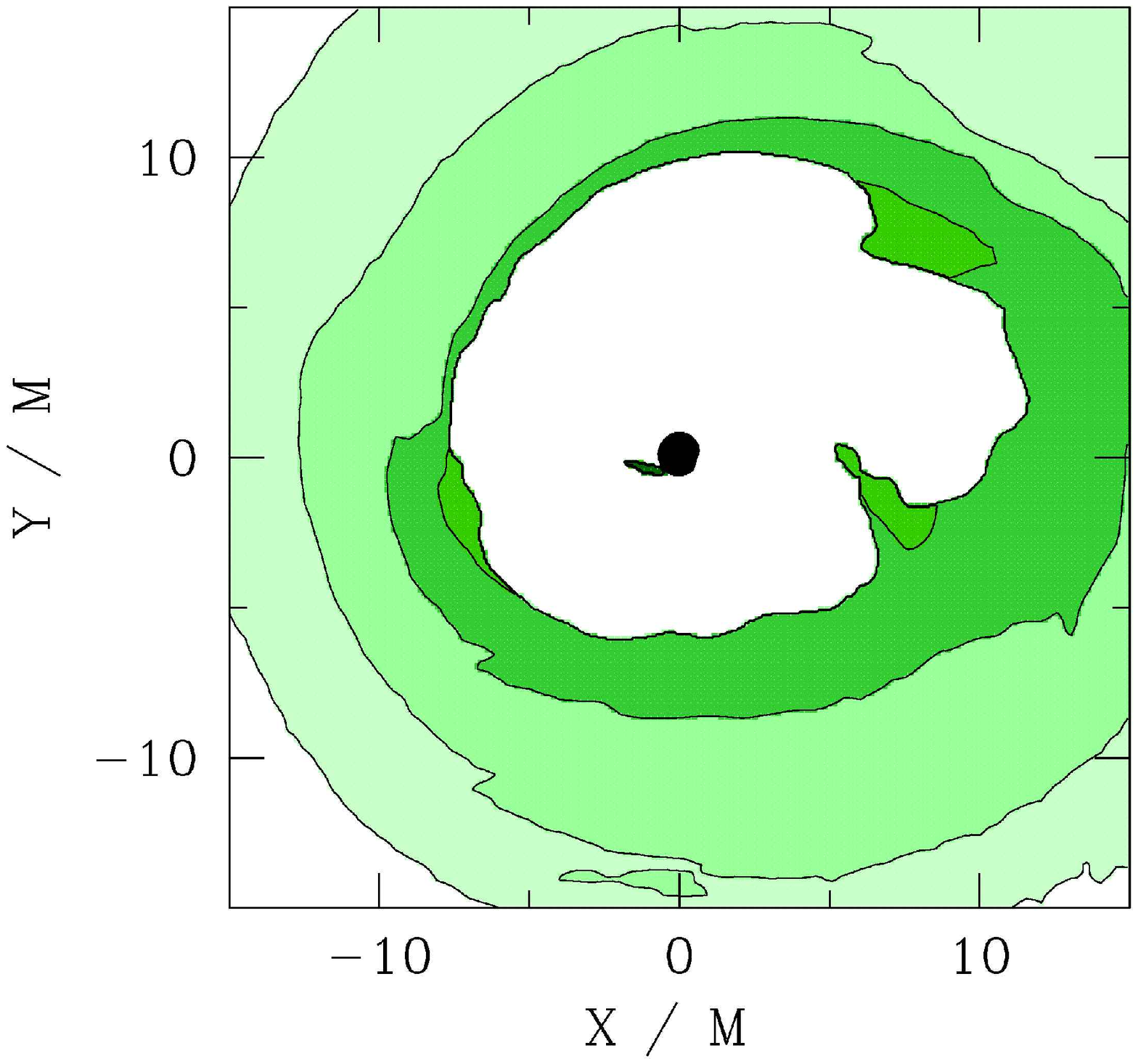}
\end{center}
\caption{Contours of entropy parameter $K$ (= $P / \rho_0^{\Gamma}$)
  at  select
  times for prograde disk with $\Gamma=5/3$.
  Contours correspond to $K/K_{0}=10^{0.1+0.3 j} \ \  (j=1,2,..
..,6)$.  Frames correspond to the beginning of the late inspiral
and merger epochs
phase at $t\approx t_{merge}-1250M$ (top),
$t\sim t_{merge}-50M$ (middle), and $t\sim t_{merge}+\tdisk$ (bottom). Regions with
density less than $\rho_0/\rho_{0,max} < 10^{-4.5}$ are left
white. Darker shading denotes higher $K$.}
  \label{fig:heating_contour}
\end{figure}

To highlight the role that shock heating plays in our
simulations, we also plot contours of $K/K_0$.  Here $K\equiv P /
\rho_0^{\Gamma}$ and $K_{0}$ is the initial value of $K$ everywhere.
The quantity $K=K(s)$, where $s$
is the specific gas entropy, remains constant in the absence of shocks; shock
heating yields $K/K_0 > 1$ (see Appendix B of \cite{etienne09}).  As
expected, we see that $K / K_0$ increases steeply near the BHs, where
the binary torque-induced spiral arms are strongest (see
Fig.~\ref{fig:heating_contour}).  Following the merger, binary torques
are no longer present and we find that $K/K_0$ is dramatically reduced
in the cavity region.  While we do find that a small region near the
remnant continues to have $K/K_0$ long after the merger (see
Fig.~\ref{fig:heating_contour}), we note that the gas in this region
is of very low density and carries a
relatively insignificant amount of thermal energy.
 
We also note that the shocks are confined to the inner region and do
not propagate into the bulk of the disk.  While it has been proposed that changes in the
potential due to mass loss and/or BH kicks following merger can give
rise to shocks throughout the disk \cite{bode07,oneill09}, we do
not observe such behavior in our simulations.  However, this is
expected, as it has been noted that a condition for shocks to form due
to mass loss is
that $\epsilon > H/R$, where $\epsilon\equiv (M_i-M_f)/M_i$ is the
fractional mass loss due to gravitational wave emission \cite{oneill09}.  Comparing the fractional mass
loss for an equal-mass merger, for which $\epsilon \approx 0.05$, to the
estimates of $H/R$ measured at the moment of binary merger (see Table~\ref{table:unmag}), we find
that the above condition is never satisfied. The criteria above for
shocks to form is derived from the condition that the radial velocity
must exceed the sound speed $c_s
\equiv (\Gamma P/\rho_0 h)^{1/2}$ near the inner edge of the disk. We have also checked this
directly and have found that the condition is never met in our
simulations - our disks are too hot, hence too geometrically thick, to
trigger this effect.


\subsection{Scaling and detectability}
In quoting values for the accretion rate
$\dot{M}_0$, we normalize by the quantity $\rho_{max} M^2 = 0.2 n_{12}M_8^2 M_{\odot}\mbox{yr}^{-1}$, which allows for
easy scaling to arbitrary disk density and binary mass. It is also possible to derive simple scaling relations for the
luminosities.  The dominant region of emission differs
for bremsstrahlung and synchrotron radiation.  Because of the stronger
dependence of synchrotron emissivity on temperature (see Appendix B of
\cite{farris10}), we find that the
synchrotron emission originates chiefly from the hot gas near the BHs,
while the majority of bremsstrahlung emission originates from the cooler, denser
gas in the bulk of the disk.
In each simulation, we find that the temperature is maximum
near the BHs, typically reaching $kT_h \approx 100 \mbox{MeV}$
at the horizon.  In the high-temperature limit ($kT
> m_e c^2$) the synchrotron emissivity given in
Appendix B of \cite{farris10} scales with temperature and density
according to
\begin{equation}
  \label{synch_emissivity}
  q_{\mbox{syn}} \propto n_h^2 T_h^3 \beta^{-1} \ .
\end{equation}
By contrast, the temperature in the bulk of the disk where the majority of
the bremsstrahlung emission originates is nonrelativistic ($k T_{disk}
< m_e c^2$) so that the bremsstrahlung emissivity scales according to
\begin{equation}
\label{brem_emissivity}
q_{ff} \propto n_{disk}^2 T_{disk}^{1/2} \ .
\end{equation}

Here $n_h$ and $T_h$ refer to the density and temperature at the
horizon, and $n_{disk}$ and $T_{disk}$ refer to the density and
temperature at $\rdisk$, and $\beta \equiv 8 \pi P/B^2$ sets the strength
of the assumed turbulent magnetic field responsible for synchrotron emission.
As in Paper I, we choose $\beta=10$ so that the magnetic field is
assumed to reach a fraction of its equipartition value consistent with
MHD simulations \cite{mckinney04}.  

Integrating Eqs.~(\ref{synch_emissivity}) and
(\ref{brem_emissivity}), we find
\begin{eqnarray}
  L_{syn} &\approx& \int dV q_{syn}  \propto n_h^2 T_h^3 \beta^{-1} M^3 \ ,\\
  L_{ff} &\approx& \int dV q_{ff} \propto   n_{disk}^2 T_{disk}^{1/2}
  R_{disk}^3 (H/R) \ ,
\end{eqnarray}
where the factor of $M^3$ comes from the volume element.

Because we ignore self-gravity in the disk, our scaling
results apply for arbitrary density. Accordingly, as the
density at the horizon is varied, its value is
simply proportional to the maximum density in the disk, $n_h \propto
n_{disk} \propto \rho_{max}$.
 
Consider the
onset of the binary decoupling (late inspiral) phase. The spiral arms through which the gas enters the disk cavity
are shock heated.  The gas in any shocked
region will be heated to $kT\sim m_B v^2$.  Since $v \lesssim c$ near
the horizon, shock heating guarantees that $k T_h \lesssim m_B c^2
\sim 10^{3} \mbox{ MeV}$, independent of the temperature in the bulk of the
disk (in fact $K T_h$ is closer to 100 MeV). Scaling for
synchrotron luminosity then simplifies to
\begin{equation}
   L_{syn} \propto \rho_{max}^2 \beta^{-1} M^3 \ .
\end{equation}

While the bremsstrahlung luminosity does
depend on $T_{disk}$, we note that for the fixed values of $q$, $l(\rin)$, and
$\rin/M$ specified in Eqs.~(\ref{q_disk}), (\ref{lin_disk}), and (\ref{rin_disk}), the
enthalpy profile $h$, and therefore $k T_{disk} = m_B (h_{disk}-1) (\Gamma-1)/\Gamma$, is uniquely specified.  We therefore regard $T_{disk}$ as a fixed
parameter, so that
\begin{equation}
 L_{ff} \propto \rho_{max}^2 M^3 \ .
\end{equation}

Using these scaling relations, along with the results of our
simulations, we estimate the average luminosity when the binary
separation is $a=10M$, the adopted onset of the late inspiral and
merger epochs. Results are given in Table~\ref{table:lum_results}.
\begin{table*}
\caption{Electromagnetic emission at beginning of late inspiral and
  merger epoch and shortly before merger}
\begin{tabular}{lllllll}
Case \ \ \ &Time \ \ \ &${}^a\dot{M}_0 / (\rho_{max} M^2)$ \ \ \ \ \ \ &${}^bL_{ff}/L_{46}$ \ \ \ \
&$L_{syn}/L_{46}$ \ \ \ \
&${}^c h \nu_{ff}$ [$(1+z)^{-1}$MeV]
 \ \ \ \ \  &${}^d h \nu_{syn}$ [$(1+z)^{-1}n_{12}^{1/2}\beta_1^{-1/2}$eV]\\
\hline
\hline
B1&$\tgw-1250M$&0.02&0.65&$1.2 \beta_1^{-1}$&0.3&2.5\\
B2&&0.003&0.16&$0.05 \beta_1^{-1}$&0.2&0.6\\
B3&&0.0005&0.003&$2 \times 10^{-5} \beta_1^{-1}$&0.08&0.01\\
B4&&0.01&0.15&$0.5 \beta_1^{-1}$&0.2&1.6\\
\hline
B1&$\tgw-50M$&0.0003&0.60&$ 0.3\beta_1^{-1}$&0.3&1.7\\
B2&&0.0005&0.14&$0.008 \beta_1^{-1}$&0.2&1.5\\
B3&&0.0001&0.002&$2.0 \times 10^{-6}\beta_1^{-1}$&0.08&1.7\\
B4&&0.0005&0.13&$0.0002 \beta_1^{-1}$&0.2&1.4\\
\hline
\end{tabular}
\vskip 12pt
\begin{minipage}{12cm}
  \raggedright 
  ${}^{a} \rho_{max}M^2 = 0.2 n_{12} M_8^2 M_{\odot}\mbox{yr}^{-1}$, $n_{12} \equiv n/10^{12}\mbox{cm}^{-3}$, $M_8
  \equiv M/10^8 M_{\odot}$.

  ${}^{b} L_{46} = 10^{46} n_{12}^2 M_8^3 \mbox{erg s}^{-1}$

  ${}^{c} h \nu_{ff} = k T_{disk}$

  ${}^{d}\beta_1\equiv
  \beta/10$, $\beta \equiv 8 \pi P /  B^2$.
\end{minipage}    
\label{table:lum_results}
\end{table*}
 
In calculating the luminosity, we have assumed that the gas is
optically thin.  We can verify that this is a good assumption by
estimating the optical depth.  Taking the dominant opacity
source to be electron scattering, we find
\begin{equation}
  \tau_{es} \approx n_h \sigma_T R \sim 0.2 n_{12} M_8
\end{equation}
where $R$ is the characteristic size of the emission region that we
have set to $R\approx 2M$.  Thus,
we see that our assumption of an optically thin gas is valid for our
canonical parameters, although it begins to break down when
we consider denser disks and/or more massive binaries.

For bremsstrahlung emission originating
at $\rdisk$, the
characteristic observed frequency of the emission is given by
\begin{equation}
  h \nu_{ff} \sim k T_{disk}/(1+z)
\end{equation}
for a source at redshift $z$.  We measure the
temperature at $\rdisk$ for each of our cases, and report
the estimated characteristic frequencies in Table~\ref{table:lum_results}.
For canonical parameters, bremsstrahlung emission will be predominantly in
$\gamma$ rays.  Based on our measured luminosities, we estimate that
the observed flux from this emission will be in the range of $\sim
10^{-15} - 10^{-14} n_{12}^2 M_8^3\mbox{erg cm}^{-2}\mbox{ s}^{-1}$ for a source at
$z=1$.  Unfortunately, it is unlikely that this emission is strong
enough to be detectable.  We note that the bremsstrahlung emission we
measure is actually dominated by emission from the bulk of the disk
rather than the heated gas near the BHs.  This makes the emission
even less likely to be detectable, as it exhibits only a small amount
of variability.

In contrast, the synchrotron emission is predominantly produced near
the BHs.  We can estimate the characteristic frequency of the synchrotron
emission by noting that Eq.~(B10) of \cite{farris10} is maximized when
$x_M \equiv 2 \nu / 3 \nu_0 \theta^2\approx 1.09$. Here, $\nu_0 \equiv
e B / 2 \pi m_e c$ is the cyclotron frequency and $\theta \equiv
kT/m_e c^2$. The corresponding observed frequency is
\begin{equation}
  h \nu_{syn} = \frac{1.09}{1+z} \frac{3 e h B}{4 \pi m_e c}
  \left( \frac{k T}{m_e c^2} \right)^2 \ .
\end{equation}
We can use this expression, along with measured values of density and temperature in the vicinity of the horizon, to estimate characteristic values of $h \nu_{syn}$. Values for each case at the moment of decoupling and shortly before merger are given in
Table~\ref{table:lum_results}, and typically fall in the infrared
range.  We estimate that in each case, the
synchrotron emission should be observable by the proposed Wide Field Infrared Survey Telescope (WFIRST) \cite{WFIRST}, and possibly by the Large
Synoptic Survey Telescope instrument (LSST) \cite{LSST}. Our
simulations follow the late stage of the inspiral in which the binary
separation decreases from $d=10M$ to merger. This corresponds to a time scale of $\delta t \sim 100 M_8$ hrs
during which the gradual decline in emission should be observed. 
\section{Discussion}
\label{sec:discussion}
In this paper we have performed a set of fully general-relativistic
simulations of BHBH binary mergers in a circumbinary disk. Our focus has
been identifying an observable electromagnetic
signal that may accompany the gravitational waves from a black
hole merger. Our simulations are exploratory only. We have restricted our attention to disklike accretion
onto equal-mass, nonspinning BH binaries, although our methods may be extended to other binary configurations.  We exploit the approximate helical Killing
symmetry to determine the binary spacetime for widely separated BHBHs. The disk we evolve in this early inspiral spacetime relaxes to near quasiequilibrium. Our late inspiral and merger
simulations begin with such a quasistationary disk. We then evolve
the field as well as the matter. This epoch corresponds to
the post disk-binary decoupling phase, terminating after merger, but before
viscosity fills in the hollow.

For each simulation, we have calculated the time-varying rest-mass
accretion rate, as well as the electromagnetic luminosity due to
bremsstrahlung and synchrotron emission.  We also derive scaling
relations for the luminosity, enabling our results to be applied to a
range of gas parameters and BH masses.  

In each case, we find evidence for a time-varying electromagnetic
signature accompanying the BHBH binary merger.  The
synchrotron emission is the most easily detectable component, and we
observe a steady decline in synchrotron luminosity throughout the
post-decoupling binary inspiral.  This change serves as a characteristic precursor of a
binary merger, and should be detectable by the proposed WFIRST and possibly by the LSST instrument.

In Paper I, we restricted our attention to Bondi-like accetion onto
merging binaries.  It is instructive to compare the electromagnetic
signatures associated with binary Bondi accretion with the signatures from disklike accretion
discussed in this paper.  
In the
binary Bondi case, there is a steady supply of gas accreting onto
the binary at all stages of the merger.  In this case, the evolution
of the luminosity is determined by the strength of shock heating near
the BHs.  As the separation decreases, the BHs orbit more
rapidly, and the shock-heated temperature of
the gas increases. This increase leads to a luminosity that increases throughout
the inspiral, then drops precipitously following the merger as the
shocks dissipate. This scenario is quite different from the case of disklike
accretion, in which
binary torques create a hollow region around the binary as well as a small amount of matter that leaks into the hollow in the form of spiral arms. Because the torques
decrease throughout the inspiral as the BH separation decreases, we
find that the accretion rate and luminosity decrease steadily over the
course of our inspiral simulations.  

We suspect, however, that this
picture will change with the addition of magnetic fields, as the magneto-rotational instability will lead to
turbulence. We intend to investigate this
behavior in future calculations, although we expect that our
results for the late inspiral and merger epochs treated here will not be significantly altered. The reason is that the time scale for turbulent viscosity to fill the hollow with gas for accretion exceeds the inspiral time scale following decoupling.

It is not possible to make a quantitative comparison of our
results with those of Bode {\it et al} \cite{bode11} due to
significant differences in our methods.  We employ disk solutions with
power-law rotation dependence \cite{chakrabarti85} and BHBH CTS metric
data that we allow to relax over a time scale of $\sim 5 t_{disk}
\approx 6000 M$
before we begin our inspiral calculations. By contrast, Bode {\it et al}
employ the constant midplane density initial data of \cite{oneill09}
and relax this profile for a period of $\sim 250M$. Our calculations also differ
significantly in that we consider bremmstrahlung radiation from the
entire disk, whereas they consider only the cavity region near the
BHs. We calculate the synchrotron emission as
well.  Nevertheless, we are able to see a qualitative agreement in the
evolution of the rest-mass accretion rate $\dot{M}_0$, as a decline in $\dot{M}_0$ is observed throughout the inspiral phase in both
calculations.

\acknowledgments
We would like to thank Z. Etienne, C. Gammie, and V. Paschilidis for useful
discussions.  We are also grateful to H. Pfeiffer
for providing CTS initial data for the BHBH spacetime metric. This paper was supported in part by NSF Grants No. PHY06-50377 and
No. PHY09-63136 as well as NASA Grants No. NNX07AG96G, and
No. NN11AE11G. B. Farris gratefully acknowledges support from NASA Earth
and Space Science Fellowship, NNX09AO64H. 

\appendix
\section{Disk initial data}
\label{disk_id_appendix}
Our formulation for an equilibrium stationary disk around a {\it single} Kerr BH follows
closely that of \cite{chakrabarti85,devilliers03}.

From the conservation of the stress-energy equation, we find 
\begin{eqnarray}
  0=T^{\beta}{}_{\alpha;\beta} &=& \frac{1}{\alpha \sqrt{\gamma}} \left(\alpha
    \sqrt{\gamma} T_{\alpha}{}^{\beta}\right)_{,\beta} -
  \Gamma^{\lambda}_{\alpha \mu} T_{\lambda}{}^{\mu}\\
&=&  \frac{1}{\alpha \sqrt{\gamma}} \left(\alpha
    \sqrt{\gamma} \rho_0 h u_{\alpha}u^{\beta}\right)_{,\beta}\\
&&  +
  P_{,\alpha} + \frac{1}{2}( g^{\mu \nu}{}_{,\alpha} \rho_0 h u_{\mu}
  u_{\nu} ) \ .
\label{divT0}
\end{eqnarray}

Since we are seeking a solution for a stationary torus in Kerr
spacetime for a single BH, we can now
impose time independence, axisymmetry, and no poloidal or radial
motion: 
\begin{eqnarray}
  \partial_t (\ldots) &=&\partial_{\phi}(\ldots) =0 \ ,\\
  u^r &=& u^{\theta} = 0 \ .
\end{eqnarray}
In Boyer-Lindquist and Kerr-Schild coordinates, these constraints imply that
$u_r = u_{\theta} = 0$.

We may now simplify Eq.~(\ref{divT0}) according to
\begin{eqnarray}
  0 &=& \frac{h_{,j}}{h} - \frac{1}{2}  u_{t}^2 (u_t^{-2})_{,j}
  -\frac{\Omega}{1-l\Omega}l_{,j} \ .
\label{difeq}
\end{eqnarray}
Here we have also assumed constant entropy, and we have
introduced the specific angular momentum $l \equiv  -u_{\phi} / u_t$.  We have also used
the fact that $u^{\mu}u_{\mu} = -1$ to show that
\begin{equation}
  u_t^{-2} = -(g^{tt}-2lg^{t\phi}+l^2g^{\phi\phi}) \ ,
\end{equation}
and have defined 
\begin{equation}
  \Omega \equiv u^{\phi}/u^t =
  (g^{t\phi}-lg^{\phi\phi})/(g^{tt}-lg^{t\phi}) \ .
  \end{equation}
We now follow \cite{chakrabarti85,devilliers03} and assume the disk has a power-law
rotation dependence, whereby $\Omega$ takes the form
\begin{equation}
  \label{omega_def}
  \Omega = \eta \lambda^{-q} \ ,
\end{equation}
where 
\begin{equation}
  \label{lambda_def}
  \lambda^2 \equiv \frac{l}{\Omega} = l \frac{g^{tt}-l g^{t
      \phi}}{g^{t\phi}-lg^{\phi\phi}} \ .
\end{equation}
Combining Eq.~(\ref{omega_def}) and Eq.~(\ref{lambda_def}), we find
that 
\begin{equation}
  \Omega = k l^{\alpha} \ ,
\end{equation}
where $\alpha\equiv q/(q-2)$ and $k=\eta^{-2/(q-2)}$.

It is now straightforward to show that Eq.~(\ref{difeq}) is satisfied by 
\begin{equation}
  h(r,\theta)= \frac{u_{t,in} f(l_{in})}{u_t(r,\theta)f(l(r,\theta))}
  \ ,
\end{equation}
where $f(l) = |1-kl^{\alpha+1}|^{1/(\alpha+1)}$.  A disk solution is
uniquely determined for fixed $\rin$, $l(\rin)$, and $q$.

While the solution described above applies to equilibrium disks in
general Kerr spacetimes, we assume a Schwarzschild geometry
for the purposes of this study.  

In the Newtonian limit, we know that $l= \Omega r^2$, so $\lambda^2 =
l/\Omega=r^2$, whereby
\begin{equation}
  \Omega \propto \lambda^{-q} \propto r^{-q} \ .
\end{equation}
Thus we can see that asymptotically, 
\begin{eqnarray}
  q=0 &\Rightarrow& \Omega \mbox{= const} \ ,\\
  q=2 &\Rightarrow& l=\mbox{const} \ ,\\
  q=1.5 &\Rightarrow& \mbox{Keplerian} \ .
\end{eqnarray}

In this paper, the following parameters are chosen to determine the initial disk
configuration (set close to Keplerian),
\begin{eqnarray}
  \label{q_disk}
  q &=& 1.6 \ ,\\
  \label{lin_disk}
  l(\rin) &=& 4.53 \ ,\\
  \label{rin_disk}
  \rin &=& 15 M \ .
\end{eqnarray}

\section{Derivation of the torque density $dT/dR$}
\label{torque_dens_appendix}
If we consider a four vector $\vec{j}$ that is not necessarily conserved, then we must generalize
Eq.~(\ref{cons_eq}) to
\begin{equation}    
  \label{torque_dens_1}
  \frac{dq}{dt} =-\mathcal{F}_H + \mathcal{F}_L + \int \sqrt{-g}
  \nabla_{\mu} j^{\mu} d^3x
\end{equation}  
If we choose to set $H$ and $L$ to be two concentric cylinders of
infinite extent, centered around the $z$ axis and of radius $R$ and
$R+\delta R$ respectively, we can rewrite Eq.~(\ref{torque_dens_1}) as
\begin{eqnarray}
  \frac{d}{dt}( q(R+\delta R) &-& q(R))
  =-\mathcal{F}(R) + \mathcal{F}(R+\delta R) \nonumber\\
  &&+ \int \sqrt{-g}
  \nabla_{\mu} j^{\mu} R \ dR \ dz \ d\phi
\end{eqnarray}
taking the limit $\delta R \rightarrow 0$, we therefore find that 
\begin{equation}
  \label{torque_dens_2}
  \frac{d}{dR}\left(\frac{dq}{dt}\right) = \frac{d\mathcal{F}}{dR} +
  \int \sqrt{-g}\nabla_{\mu} j^{\mu} R \ dz \ d\phi \ .
\end{equation}
We can consider the specific case where $j^{\mu}=T^{\mu}{}_{\nu}
\phi^{\nu}$, and $\phi^{\nu} \equiv (\partial_{\phi})^{\mu}=(0,-y,x,0)$, so that Eq.~(\ref{torque_dens_2}) becomes 
\begin{eqnarray}
  \label{torque_dens_3}
  \frac{d}{dR}\left(\frac{dJ}{dt}\right) &\equiv& \frac{dT_{tot}}{dR} \nonumber\\
  &=& \frac{d\mathcal{F}^{(J)}}{dR} +
  \int \sqrt{-g}T^{\mu\nu}\nabla_{\mu} \phi_{\nu} R dz d\phi \ .
\end{eqnarray}
We interpret the first term on the right-hand side of Eq.~(\ref{torque_dens_3}) as arising from the net outflow of angular momentum carried by matter across the surfaces at $R$ and $R+\delta R$, while the
second term is the torque due to the gravitational field.  Because we
are most interested in the torque from the gravitational field of the binary, we define
\begin{equation}
  \label{torque_dens_4}
  \frac{dT}{dR} \equiv \frac{dT_{tot}}{dR} -
  \frac{d\mathcal{F}^{(J)}}{dR} =   \int \sqrt{-g}T^{\mu\nu}\nabla_{\mu} \phi_{\nu} R \ dz \ d\phi \ .
\end{equation}
Note that in an axisymmetric spacetime in which $\phi_{\nu}$ is a Killing vector
field, $T^{\mu \nu} \nabla_{\mu }\phi_{\nu} = T^{\mu \nu} \nabla_{(\mu
}\phi_{\nu)} = 0$, hence $dT/dR=0$ as expected.

In order to compute Eq.~(\ref{torque_dens_4}) numerically, it is convenient to
transform the expression into Cartesian coordinates.  Note that
\begin{eqnarray}
\label{covariant_deriv}
  T^{\mu \nu} \nabla_{\mu} \phi_{\nu} &=& T^{\mu}{}_{\nu} \nabla_{\mu}
  \phi^{\nu} \nonumber\\
 &=&  T^{\mu}{}_{\nu}(\phi^{\nu}_{,\mu} + \Gamma^{\nu}{}_{\sigma \mu}
  \phi^{\sigma})\nonumber\\
  &=& T^{\mu}{}_{\nu}\phi^{\nu}_{,\mu}+ \frac{1}{2} T^{\mu \nu} g_{\mu
    \nu, \sigma} \phi^{\sigma}\nonumber\\
  &=& -T^{\mu}{}_{x}\ y_{,\mu}
  +T^{\mu}{}_{y}\ x_{,\mu}\nonumber\\
  &&+\frac{1}{2}T^{\mu \nu}\left( -g_{\mu \nu,x}y
  +g_{\mu \nu,y}x\right) \nonumber\\
  &=& -T^{y}{}_{x}+T^{x}{}_{y}\nonumber\\
  &&-\frac{1}{2}yT^{\mu \nu}g_{\mu \nu,x}+\frac{1}{2}xT^{\mu \nu} g_{\mu \nu,y}
\end{eqnarray}
Inserting Eq.~(\ref{covariant_deriv}) into Eq.~(\ref{torque_dens_4}),
we find
\begin{eqnarray}
   \label{torque_dens_5}
\frac{dT}{dR} &=& \int R\  d\phi \ dz  \sqrt{-g}\left( -T^{y}{}_{x} + T^x{}_y\right)\\
&&+\frac{1}{2}\int  R\  d\phi \ dz  \sqrt{-g}\left( -yT^{\mu \nu}g_{\mu
    \nu,x}+xT^{\mu \nu} g_{\mu \nu,y}\right) \ . \nonumber
\end{eqnarray}
Eq.~(\ref{torque_dens_5}) is integrated numerically at a number of
different radii, so that we can compute profiles of $dT/dR$.
\subsection{Newtonian limit}
We can check that Eq.~(\ref{torque_dens_5}) reduces to the correct
expression in the Newtonian limit.  If we let $T^{00} \approx \rho_0$, $|T^{0i}/T^{00}| \ll 1$, and $|T^{ij}/T^{00}| \ll 1$, we find
\begin{eqnarray}
\label{newtonian}
\frac{dT}{dR} &\approx& \frac{1}{2} \int R \ d \phi \ dz \ T^{00}(-y
g_{00,x}+xg_{00,y})\nonumber\\
&=& \frac{1}{2} \int R \ d \phi \ dz \ T^{00} g_{00,\phi}\nonumber\\
&=&- \int R \ d \phi \ dz \ \rho_0  \Phi_{,\phi}\nonumber\\
&=& - 2 \pi R \left< \Sigma \Phi_{,\phi} \right>
\end{eqnarray}
where angled brackets indicate angle averaging, and $\Phi$ is the
Newtonian gravitational potential.  Eq.~(\ref{newtonian}) matches the expression
given in Eq.~(14) of \cite{macfadyen08}.

\bibliography{bbh_disk_paper}
\end{document}